\documentclass[a4paper]{article}\usepackage[]{graphicx}\usepackage[]{color}
\makeatletter
\def\maxwidth{ %
  \ifdim\Gin@nat@width>\linewidth
    \linewidth
  \else
    \Gin@nat@width
  \fi
}
\makeatother

\definecolor{fgcolor}{rgb}{0.345, 0.345, 0.345}
\newcommand{\hlnum}[1]{\textcolor[rgb]{0.686,0.059,0.569}{#1}}%
\newcommand{\hlstr}[1]{\textcolor[rgb]{0.192,0.494,0.8}{#1}}%
\newcommand{\hlopt}[1]{\textcolor[rgb]{0,0,0}{#1}}%
\newcommand{\hlstd}[1]{\textcolor[rgb]{0.345,0.345,0.345}{#1}}%
\newcommand{\hlkwb}[1]{\textcolor[rgb]{0.69,0.353,0.396}{#1}}%
\newcommand{\hlkwc}[1]{\textcolor[rgb]{0.333,0.667,0.333}{#1}}%
\newcommand{\hlkwd}[1]{\textcolor[rgb]{0.737,0.353,0.396}{\textbf{#1}}}%

\usepackage{framed}
\makeatletter
\newenvironment{kframe}{%
 \def\at@end@of@kframe{}%
 \ifinner\ifhmode%
  \def\at@end@of@kframe{\end{minipage}}%
  \begin{minipage}{\columnwidth}%
 \fi\fi%
 \def\FrameCommand##1{\hskip\@totalleftmargin \hskip-\fboxsep
 \colorbox{shadecolor}{##1}\hskip-\fboxsep
     \hskip-\linewidth \hskip-\@totalleftmargin \hskip\columnwidth}%
 \MakeFramed {\advance\hsize-\width
   \@totalleftmargin\z@ \linewidth\hsize
   \@setminipage}}%
 {\par\unskip\endMakeFramed%
 \at@end@of@kframe}
\makeatother

\definecolor{shadecolor}{rgb}{.97, .97, .97}
\definecolor{messagecolor}{rgb}{0, 0, 0}
\definecolor{warningcolor}{rgb}{1, 0, 1}
\definecolor{errorcolor}{rgb}{1, 0, 0}
\newenvironment{knitrout}{}{} 

\usepackage{alltt}
\usepackage[margin=1.3in]{geometry}
\usepackage[utf8]{inputenc}
\usepackage[T1]{fontenc}
\usepackage{amsmath,amssymb,array}
\usepackage{booktabs}
\usepackage{hyperref}

\RequirePackage[sectionbib,round]{natbib}

\newcommand{\CRANpkg}{\textbf}
\newcommand{\pkg}{\textbf}
\newcommand{\code}{\texttt}
\newcommand{\dfn}{\emph}
\newcommand{\samp}{\texttt}

\IfFileExists{upquote.sty}{\usepackage{upquote}}{}
\begin{document}




\title{\pkg{broom}: An {R} Package for Converting Statistical Analysis Objects Into Tidy Data Frames}
\author{David Robinson}
\date{}

\maketitle

\begin{abstract}
The concept of "tidy data" offers a powerful framework for structuring data to ease manipulation, modeling and visualization. However, most R functions, both those built-in and those found in third-party packages, produce output that is not tidy, and that is therefore difficult to reshape, recombine, and otherwise manipulate. Here I introduce the \CRANpkg{broom} package, which turns the output of model objects into tidy data frames that are suited to further analysis, manipulation, and visualization with input-tidy tools. \pkg{broom} defines the \code{tidy}, \code{augment}, and \code{glance} generics, which arrange a model into three levels of tidy output respectively: the component level, the observation level, and the model level. I provide examples to demonstrate how these generics work with tidy tools to allow analysis and modeling of data that is divided into subsets, to recombine results from bootstrap replicates, and to perform simulations that investigate the effect of varying input parameters.
\end{abstract}

\section{Introduction}

A common observation is that more of the data scientist's time is occupied with data cleaning, manipulation, and "munging" than it is with actual statistical modeling \citep{Rahm:2000tl,Dasu:2003ul}. Thus, the development of tools for manipulating and transforming data is necessary for efficient and effective data analysis. One important choice for a data scientist working in R is how data should be structured, particularly the choice of dividing observations across rows, columns, and multiple tables.

The concept of "tidy data," introduced by \citet{Wickham:2014vp}, offers a set of guidelines for organizing data in order to facilitate statistical analysis and visualization. In short, data can be described as "tidy" if it is represented in a table following three rules:

\begin{itemize}
\item Each variable forms one column
\item Each observation forms one row
\item Each type of observational unit forms one table
\end{itemize}

This framework makes it easy for analysts to reshape, combine, group and otherwise manipulate data. Packages such as \CRANpkg{ggplot2}, \CRANpkg{dplyr}, and many built-in R modeling and plotting functions require the input to be in a tidy form, so keeping the data in this form allows multiple tools to be used in sequence in a seamless analysis pipeline \citep{package:ggplot2,package:dplyr}.

Tools are classified as "messy-output" if their output does not fit into this framework. Unfortunately, the majority of R modeling tools, both from the built-in \pkg{stats} package and those in common third party packages, are messy-output. This means the data analyst must tidy not only the original data, but the results at each intermediate stage of an analysis. \citet{Wickham:2014vp} describes this problem succinctly:

\begin{quote}
While model inputs usually require tidy inputs, such attention to detail doesn't carry over to model outputs. Outputs such as predictions and estimated coefficients aren't always tidy. For example, in R, the default representation of model coefficients is not tidy because it does not have an explicit variable that records the variable name for each estimate, they are instead recorded as row names. In R, row names must be unique, so combining coefficients from many models (e.g., from bootstrap resamples, or subgroups) requires workarounds to avoid losing important information. This knocks you out of the flow of analysis and makes it harder to combine the results from multiple models. I'm not currently aware of any packages that resolve this problem.
\end{quote}

The \pkg{broom} package is an attempt to solve this issue, by bridging the gap from untidy outputs of predictions and estimations to create tidy data that is easy to manipulate with standard tools. It centers around three S3 methods, \code{tidy}, \code{augment}, and \code{glance}, that each take an object produced by R statistical functions (such as \code{lm}, \code{t.test}, and \code{nls}) or by popular third-party packages (such as \CRANpkg{glmnet}, \CRANpkg{survival}, \CRANpkg{lme4}, and \CRANpkg{multcomp}) and convert it into a tidy data frame without rownames \citep{Friedman:2010wm,package:survival,package:lme4,package:multcomp}. These outputs can then be used with input-tidy tools such as \pkg{dplyr} or \pkg{ggplot2}, or downstream statistical tests.

\pkg{broom} should be distinguished from packages such as \CRANpkg{reshape2} and \CRANpkg{tidyr}, which rearrange and reshape data frames into different forms \citep{package:reshape2,package:tidyr}. Those packages perform essential tasks in tidy data analysis but focus on manipulating data frames in one specific format into another. In contrast, \pkg{broom} is designed to take data that is \emph{not} in a data frame (sometimes not anywhere close) and convert it to a tidy data frame.

\section{Three methods of tidying objects}

As a simple demonstration of tidying a statistical object, consider a linear regression on the built-in \code{mtcars} dataset, predicting the fuel efficiency of cars (\code{mpg}, measured in miles-per-gallon) based on the weight of the cars (\code{wt}, measured in thousands of pounds) and the speed/acceleration (\code{qsec}, the time in seconds to drive a quarter of a mile).

\begin{knitrout}
\definecolor{shadecolor}{rgb}{0.969, 0.969, 0.969}\color{fgcolor}\begin{kframe}
\begin{alltt}
\hlstd{fit} \hlkwb{<-} \hlkwd{lm}\hlstd{(mpg} \hlopt{~} \hlstd{wt} \hlopt{+} \hlstd{qsec,} \hlkwc{data} \hlstd{= mtcars)}
\hlkwd{summary}\hlstd{(fit)}
\end{alltt}
\begin{verbatim}
## 
## Call:
## lm(formula = mpg ~ wt + qsec, data = mtcars)
## 
## Residuals:
##     Min      1Q  Median      3Q     Max 
## -4.3962 -2.1431 -0.2129  1.4915  5.7486 
## 
## Coefficients:
##             Estimate Std. Error t value Pr(>|t|)    
## (Intercept)  19.7462     5.2521   3.760 0.000765 ***
## wt           -5.0480     0.4840 -10.430 2.52e-11 ***
## qsec          0.9292     0.2650   3.506 0.001500 ** 
## ---
## Signif. codes:  0 '***' 0.001 '**' 0.01 '*' 0.05 '.' 0.1 ' ' 1
## 
## Residual standard error: 2.596 on 29 degrees of freedom
## Multiple R-squared:  0.8264,	Adjusted R-squared:  0.8144 
## F-statistic: 69.03 on 2 and 29 DF,  p-value: 9.395e-12
\end{verbatim}
\end{kframe}
\end{knitrout}

The summary of this regression shows that it contains coefficient-level information, including the estimate, standard error, and p-value, about each of the intercept, \code{wt}, and \code{qsec} terms. The \code{fit} object \emph{also} contains observation-level information, such as the residuals (accessed by \code{residuals(fit)}) and fitted values (accessed by \code{fitted(fit)}). Finally, it contains model-level information in the form of $R^2$, adjusted $R^2$, an F statistic, and a p-value for the whole dataset. As previously observed by \citet{Wickham:2007}, values computed at each of these three levels have different dimensionalities and observations: there is no natural way to combine a calculation of $R^2$ with the estimates of coefficient values, or with a vector of residuals, in a single data frame. In the tidy data terminology, each level forms a separate "observational unit" and therefore deserves its own table. To generate these three separate tidy data frames, \pkg{broom} provides three S3 methods that do three distinct kinds of tidying: \code{tidy}, \code{augment} and \code{glance}.

\code{tidy} constructs a data frame that summarizes the model's statistical components, which we refer to as the \dfn{component level}. In a regression such as the above it may refer to coefficient estimates, p-values, and standard errors for each term in a regression. The \code{tidy} generic is flexible- in other models it could represent per-cluster information in clustering applications, or per-test information for multiple comparison functions.

\begin{knitrout}
\definecolor{shadecolor}{rgb}{0.969, 0.969, 0.969}\color{fgcolor}\begin{kframe}
\begin{alltt}
\hlkwd{library}\hlstd{(broom)}
\hlkwd{tidy}\hlstd{(fit)}
\end{alltt}
\begin{verbatim}
##          term  estimate std.error  statistic      p.value
## 1 (Intercept) 19.746223 5.2520617   3.759709 7.650466e-04
## 2          wt -5.047982 0.4839974 -10.429771 2.518948e-11
## 3        qsec  0.929198 0.2650173   3.506179 1.499883e-03
\end{verbatim}
\end{kframe}
\end{knitrout}

\code{augment} add columns to the original data that was modeled, thus working at the \dfn{observation level}. This includes predictions, residuals and prediction standard errors in a regression, and can represent cluster assignments or classifications in other applications. By convention, each new column starts with \samp{.} to ensure it does not conflict with existing columns. To ensure that the output is tidy and can be recombined, rownames in the original data, if present, are added as a column called \code{.rownames}.

\begin{knitrout}
\definecolor{shadecolor}{rgb}{0.969, 0.969, 0.969}\color{fgcolor}\begin{kframe}
\begin{alltt}
\hlkwd{head}\hlstd{(}\hlkwd{augment}\hlstd{(fit))}
\end{alltt}
\begin{verbatim}
##           .rownames  mpg    wt  qsec  .fitted   .se.fit      .resid
## 1         Mazda RX4 21.0 2.620 16.46 21.81511 0.6832424 -0.81510855
## 2     Mazda RX4 Wag 21.0 2.875 17.02 21.04822 0.5468271 -0.04822401
## 3        Datsun 710 22.8 2.320 18.61 25.32728 0.6397681 -2.52727880
## 4    Hornet 4 Drive 21.4 3.215 19.44 21.58057 0.6231436 -0.18056924
## 5 Hornet Sportabout 18.7 3.440 17.02 18.19611 0.5120709  0.50388581
## 6           Valiant 18.1 3.460 20.22 21.06859 0.8032106 -2.96858808
##         .hat   .sigma      .cooksd  .std.resid
## 1 0.06925986 2.637300 2.627038e-03 -0.32543724
## 2 0.04436414 2.642112 5.587076e-06 -0.01900129
## 3 0.06072636 2.595763 2.174253e-02 -1.00443793
## 4 0.05761138 2.641895 1.046036e-04 -0.07164647
## 5 0.03890382 2.640343 5.288512e-04  0.19797699
## 6 0.09571739 2.575422 5.101445e-02 -1.20244126
\end{verbatim}
\end{kframe}
\end{knitrout}

Finally, \code{glance} constructs a concise one-row summary of the \dfn{model level} values. In a regression this typically contains values such as $R^2$, adjusted $R^2$, residual standard error, Akaike Information Criterion (AIC), or deviance. In other applications it can include calculations such as cross validation accuracy or prediction error that are computed once for the entire model.

\begin{knitrout}
\definecolor{shadecolor}{rgb}{0.969, 0.969, 0.969}\color{fgcolor}\begin{kframe}
\begin{alltt}
\hlkwd{glance}\hlstd{(fit)}
\end{alltt}
\begin{verbatim}
##   r.squared adj.r.squared    sigma statistic      p.value df    logLik
## 1 0.8264161     0.8144448 2.596175  69.03311 9.394765e-12  3 -74.36025
##        AIC      BIC deviance df.residual
## 1 156.7205 162.5834 195.4636          29
\end{verbatim}
\end{kframe}
\end{knitrout}

These three methods appear across many analyses; indeed, the fact that these three levels must be combined into a single S3 object is a common reason that model outputs are not tidy. Importantly, some model objects may have only one or two of these methods defined. (For example, there is no sense in which a Student's T test or correlation test generates information about each observation, and therefore no \code{augment} method exists). Table \ref{tab:tidiers} shows the tidying methods that are implemented for each statistical object that \pkg{broom} can tidy.

\begin{table}[ht]
\centering

\begin{tabular}{|c|p{7cm}|ccc|}
  \hline
package & class & tidy & augment & glance \\ 
  \hline
\pkg{base} & data.frame & X &   & X \\ 
   \cline{2-5}
   & table & X &  &  \\ 
   \hline
\pkg{stats} & anova, aov, density, ftable, manova, pairwise.htest, spec, ts, TukeyHSD & X &  &  \\ 
   \cline{2-5}
   & kmeans, lm, nls & X & X & X \\ 
   \cline{2-5}
   & smooth.spline &  & X & X \\ 
   \cline{2-5}
   & Arima, htest & X &  & X \\ 
   \cline{2-5}
   & glm &  &  & X \\ 
   \hline
\pkg{glmnet} & cv.glmnet, glmnet & X &  & X \\ 
   \hline
\pkg{lfe} & felm & X & X & X \\ 
   \hline
\pkg{lme4} & mer, merMod & X & X & X \\ 
   \hline
\pkg{maps} & map & X &  &  \\ 
   \hline
\pkg{MASS} & ridgelm & X &  & X \\ 
   \hline
\pkg{multcomp} & cld, confint.glht, glht, summary.glht & X &  &  \\ 
   \hline
\pkg{sp} & Line, Lines, Polygon, Polygons, SpatialLinesDataFrame, SpatialPolygons, SpatialPolygonsDataFrame & X &  &  \\ 
   \hline
\pkg{survival} & aareg, cch, pyears, survexp, survfit & X &  & X \\ 
   \cline{2-5}
   & coxph, survreg & X & X & X \\ 
   \hline
\pkg{zoo} & zoo & X &  &  \\ 
   \hline
\end{tabular}

\caption{The statistical objects for which \code{tidy}, \code{augment}, and \code{glance} methods are implemented in \pkg{broom} (version 0.3.4). \label{tab:tidiers}}

\end{table}

\subsection{Common attributes of messy outputs}

Extracting these three levels of tidied contents from a statistical object is not a trivial process, since they are often stored in a messy format that is not conducive to further analysis. The process of tidying a model output must be tailored to each object, since each model object is messy in its own way, but one can recognize some common features of messy-output models. Many are the same issues described in \citet{Wickham:2014vp} as features of messy datasets, such as having variables stored in column names. The following tendencies, however, are more specific to model outputs:

\begin{itemize}
\item \emph{Relevant information is stored in rownames.} Examples include the coefficient names in a regression coefficient matrix or ANOVA table. Since R does not allow row names to be duplicated, this prevents one from combining the results of multiple analyses or bootstrap replications.
\item \emph{Column names are inconsistent and inconvenient.} For instance, p-values for each coefficient produced by the \code{summary.lm} function are stored in a column named \code{Pr(>|t|)}. Besides being incomparable to other model outputs that use \code{p.value}, \code{pvalue}, or just \code{p}, this column name is difficult to work with due to the use of punctuation; for instance, it cannot easily be extracted using \samp{\$} or passed to \code{aes} in a \pkg{ggplot2} call.
\item \emph{Information is computed by downstream functions.} Many models need to be run through additional processing steps, often the \texttt{summary} method, to produce the statistical information desired. The steps required can be inconsistent even between similar models. For example, the \code{anova} function produces an ANOVA table immediately, while an object from \code{aov} must be run through \code{summary.aov} to produce the table.
\item \emph{Information is printed rather than returned.} Some packages and functions place relevant calculations into the \code{print} method, where they are displayed using \code{cat}. An example is the calculation of a p-value from an F-statistic in \code{print.summary.lm}. This requires either examining the source of the \code{print} method to extract the code of interest or capturing and parsing the printed output.
\item \emph{Vectors are stored separately rather than combined in a table.} For example, residuals and fitted values are both returned by many model outputs, but are accessed with the \texttt{residuals} and \texttt{fitted} generics. In other objects multiple vectors of the same length are included as separate elements in a named list. This requires recombining these vectors into a data frame to use them with input-tidy tools.
\end{itemize}

Each of these obstacles can be individually overcome by the knowledgeable programmer, but in combination they serve as a massive inconvenience. Time spent reforming these model outputs into the desired structure breaks the natural flow of analysis and takes attention away from examining and questioning the data. They further invite inconsistency, where different analysts will approach the same task in very different ways, which makes it time-consuming to understand and adapt shared code. Defining a standard "tidy form" of any model output, and collecting tools for constructing such a form from an R object, makes analyses easy, efficient, and consistent.

\section{Case studies}

Here I show how \pkg{broom} can be widely useful across data science applications. I give three examples: combining regressions performed on subgroups of data, a demonstration of bootstrapping, and a simulation of k-means clustering. Each example highlights some of the advantages of keeping model outputs tidy.

These examples assume familiarity with the \pkg{dplyr} and \pkg{ggplot2} packages, since they are powerful implementations of tidy tools (though it is certainly possible to take advantage of \pkg{broom} without these packages). Note that in contrast to these packages, \pkg{broom} functions take up only a small part of these examples, which is by design. \pkg{broom} is meant to serve as a simple bridge between the modeling tools provided by R and the downstream analyses enabled by tidy tools, requiring minimal experience with or configuration of the package.

\subsection{Split-apply-combine using \pkg{broom} and \pkg{dplyr}}

\label{sec:split_apply_combine}

The "split-apply-combine" pattern, where a dataset is broken down into subgroups for some analysis to be performed before being recombined, is a common task in data analysis \citep{Wickham:2011}. \pkg{broom} eases the application of this pattern to a wide range of problems, since it converts many analysis outputs to a consistently-structured data frame that can be recombined. Here we demonstrate examples of using \pkg{broom} to perform the split-apply-combine pattern, first in a simple regression on the \code{mtcars} dataset, then in a more complex analysis of baseball statistics.

We earlier showed an example of tidying a regression on the \code{mtcars} dataset.

\begin{knitrout}
\definecolor{shadecolor}{rgb}{0.969, 0.969, 0.969}\color{fgcolor}\begin{kframe}
\begin{alltt}
\hlstd{regression} \hlkwb{<-} \hlkwd{lm}\hlstd{(mpg} \hlopt{~} \hlstd{wt} \hlopt{+} \hlstd{qsec,} \hlkwc{data} \hlstd{= mtcars)}
\hlkwd{tidy}\hlstd{(regression,} \hlkwc{conf.int} \hlstd{=} \hlnum{TRUE}\hlstd{)}
\end{alltt}
\begin{verbatim}
##          term  estimate std.error  statistic      p.value   conf.low
## 1 (Intercept) 19.746223 5.2520617   3.759709 7.650466e-04  9.0045503
## 2          wt -5.047982 0.4839974 -10.429771 2.518948e-11 -6.0378678
## 3        qsec  0.929198 0.2650173   3.506179 1.499883e-03  0.3871768
##   conf.high
## 1 30.487895
## 2 -4.058096
## 3  1.471219
\end{verbatim}
\end{kframe}
\end{knitrout}

Suppose we wished to perform this analysis separately for each subset of the data: specifically, once for cars with automatic transmissions and once for cars with manual transmissions. \pkg{dplyr}'s \code{group\_by} and \code{do} functions provide a straightforward way to perform an analysis on each subgroup, but only if the output of each individual analysis takes the form of a data frame. \pkg{broom} accomplishes this goal by converting each model object into a tidy format.

\begin{knitrout}
\definecolor{shadecolor}{rgb}{0.969, 0.969, 0.969}\color{fgcolor}\begin{kframe}
\begin{alltt}
\hlkwd{library}\hlstd{(dplyr)}
\hlstd{regressions} \hlkwb{<-} \hlstd{mtcars} \hlopt{%>%} \hlkwd{group_by}\hlstd{(am)} \hlopt{%>%}
    \hlkwd{do}\hlstd{(}\hlkwd{tidy}\hlstd{(}\hlkwd{lm}\hlstd{(mpg} \hlopt{~} \hlstd{wt} \hlopt{+} \hlstd{qsec,} \hlkwc{data} \hlstd{= .),} \hlkwc{conf.int} \hlstd{=} \hlnum{TRUE}\hlstd{))}
\hlstd{regressions}
\end{alltt}
\begin{verbatim}
## Source: local data frame [6 x 8]
## Groups: am
## 
##   am        term   estimate  std.error statistic      p.value    conf.low
## 1  0 (Intercept) 11.2489412  6.7148019  1.675245 0.1133158633 -2.98580299
## 2  0          wt -2.9962762  0.6635548 -4.515491 0.0003520832 -4.40294960
## 3  0        qsec  0.9454396  0.2945500  3.209776 0.0054642663  0.32102149
## 4  1 (Intercept) 20.1753989 11.1990599  1.801526 0.1017988425 -4.77766163
## 5  1          wt -6.7543597  1.4305934 -4.721369 0.0008147494 -9.94192037
## 6  1        qsec  1.1809718  0.4924515  2.398148 0.0374338987  0.08372136
## Variables not shown: conf.high (dbl)
\end{verbatim}
\end{kframe}
\end{knitrout}

This shows the estimated coefficients of regressions within automatic and manual cars, recombined. This makes it easy to perform downstream visualizations and analyses on these regression results. For example, we could use \pkg{ggplot2} to create a coefficient plot that distinguished between the two types of transmission:

\begin{knitrout}
\definecolor{shadecolor}{rgb}{0.969, 0.969, 0.969}\color{fgcolor}\begin{kframe}
\begin{alltt}
\hlkwd{library}\hlstd{(ggplot2)}
\hlstd{regressions} \hlopt{%>%} \hlkwd{filter}\hlstd{(term} \hlopt{!=} \hlstr{"(Intercept)"}\hlstd{)} \hlopt{%>%}
    \hlkwd{ggplot}\hlstd{(}\hlkwd{aes}\hlstd{(}\hlkwc{x} \hlstd{= estimate,} \hlkwc{y} \hlstd{= term,} \hlkwc{color} \hlstd{=} \hlkwd{factor}\hlstd{(am),} \hlkwc{group} \hlstd{= am))} \hlopt{+}
    \hlkwd{geom_point}\hlstd{()} \hlopt{+} \hlkwd{geom_errorbarh}\hlstd{(}\hlkwd{aes}\hlstd{(}\hlkwc{xmin} \hlstd{= conf.low,} \hlkwc{xmax} \hlstd{= conf.high))} \hlopt{+}
    \hlkwd{geom_vline}\hlstd{()}
\end{alltt}
\end{kframe}
\includegraphics[width=3in,height=3in]{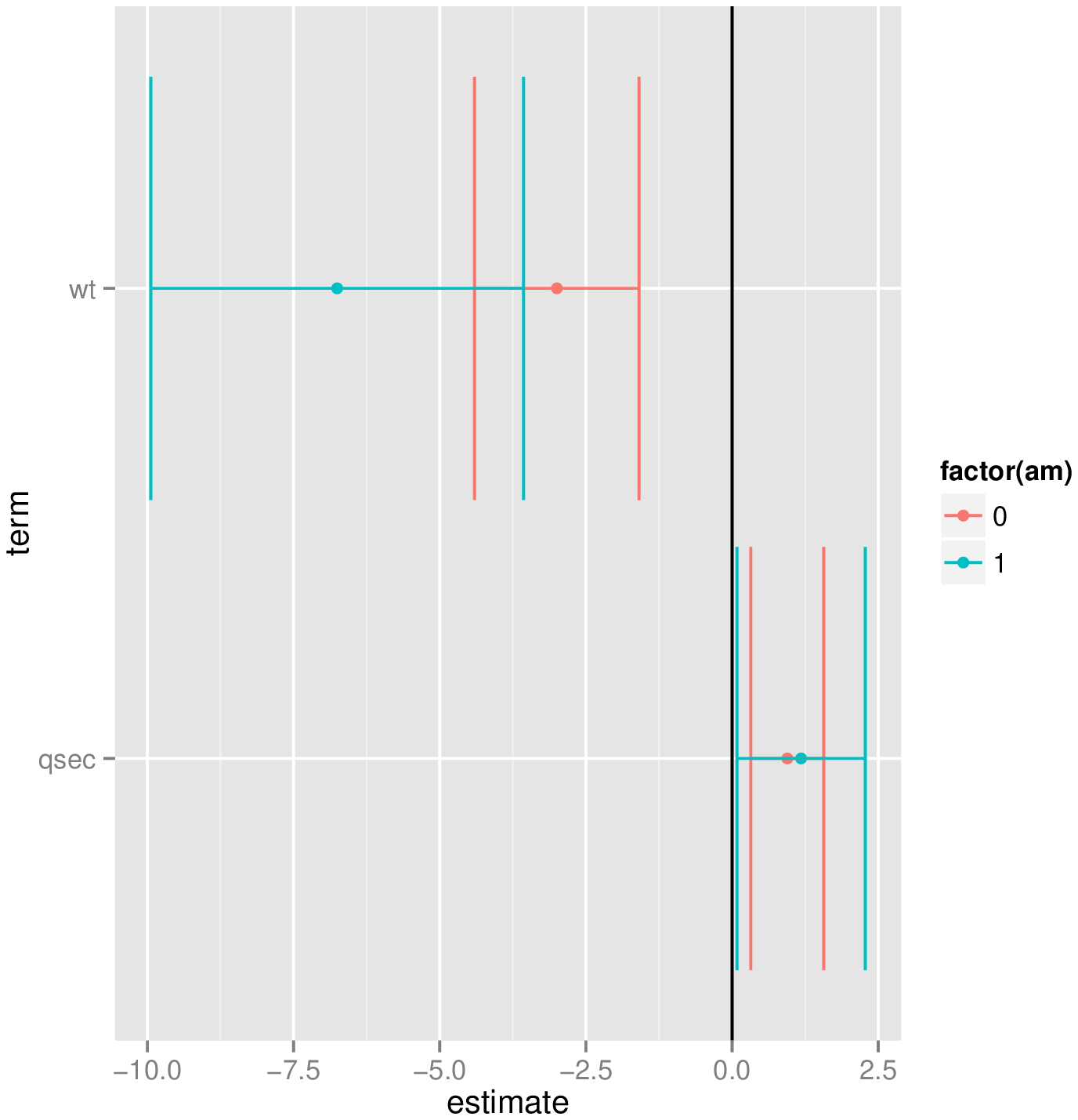} 

\end{knitrout}

The \textbf{do} function also allows one to construct regressions once and save them as a column in a rowwise table, so that we can perform multiple operations on them.

\begin{knitrout}
\definecolor{shadecolor}{rgb}{0.969, 0.969, 0.969}\color{fgcolor}\begin{kframe}
\begin{alltt}
\hlstd{regressions} \hlkwb{<-} \hlstd{mtcars} \hlopt{%>%} \hlkwd{group_by}\hlstd{(am)} \hlopt{%>%}
    \hlkwd{do}\hlstd{(}\hlkwc{mod} \hlstd{=} \hlkwd{lm}\hlstd{(mpg} \hlopt{~} \hlstd{wt} \hlopt{+} \hlstd{qsec, .))}
\hlstd{regressions}
\end{alltt}
\begin{verbatim}
## Source: local data frame [2 x 2]
## Groups: <by row>
## 
##   am     mod
## 1  0 <S3:lm>
## 2  1 <S3:lm>
\end{verbatim}
\end{kframe}
\end{knitrout}

This is useful if we want to tidy, augment and glance each regression separately. Indeed, \pkg{broom} includes a shortcut for this purpose, where \code{\%>\% tidy(`column`)} on a rowwise table will perform a tidying operation on each object in that column:

\begin{knitrout}
\definecolor{shadecolor}{rgb}{0.969, 0.969, 0.969}\color{fgcolor}\begin{kframe}
\begin{alltt}
\hlstd{regressions} \hlopt{%>%} \hlkwd{tidy}\hlstd{(mod,} \hlkwc{conf.int} \hlstd{=} \hlnum{TRUE}\hlstd{)}
\end{alltt}
\begin{verbatim}
## Source: local data frame [6 x 8]
## Groups: am
## 
##   am        term   estimate  std.error statistic      p.value    conf.low
## 1  0 (Intercept) 11.2489412  6.7148019  1.675245 0.1133158633 -2.98580299
## 2  0          wt -2.9962762  0.6635548 -4.515491 0.0003520832 -4.40294960
## 3  0        qsec  0.9454396  0.2945500  3.209776 0.0054642663  0.32102149
## 4  1 (Intercept) 20.1753989 11.1990599  1.801526 0.1017988425 -4.77766163
## 5  1          wt -6.7543597  1.4305934 -4.721369 0.0008147494 -9.94192037
## 6  1        qsec  1.1809718  0.4924515  2.398148 0.0374338987  0.08372136
## Variables not shown: conf.high (dbl)
\end{verbatim}
\begin{alltt}
\hlstd{regressions} \hlopt{%>%} \hlkwd{glance}\hlstd{(mod)}
\end{alltt}
\begin{verbatim}
## Source: local data frame [2 x 12]
## Groups: am
## 
##   am r.squared adj.r.squared    sigma statistic      p.value df    logLik
## 1  0 0.7501667     0.7189375 2.032590  24.02135 1.517760e-05  3 -38.80416
## 2  1 0.8896117     0.8675340 2.244353  40.29465 1.639138e-05  3 -27.25026
## Variables not shown: AIC (dbl), BIC (dbl), deviance (dbl), df.residual
##   (int)
\end{verbatim}
\end{kframe}
\end{knitrout}

This split-apply-combine pattern appears in a variety of contexts. One might perform analyses for each gene in an organism, within each county or region in a country, or for each stock in a financial dataset. Here I consider another example, where we investigate how baseball players' batting averages are reflected in their salaries, using data from the \CRANpkg{Lahman} package \citep{package:Lahman}.

\begin{knitrout}
\definecolor{shadecolor}{rgb}{0.969, 0.969, 0.969}\color{fgcolor}\begin{kframe}
\begin{alltt}
\hlkwd{library}\hlstd{(Lahman)}
\hlstd{merged} \hlkwb{<-} \hlstd{Batting} \hlopt{%>%} \hlkwd{tbl_df}\hlstd{()} \hlopt{%>%}
    \hlkwd{inner_join}\hlstd{(Salaries)} \hlopt{%>%}
    \hlkwd{mutate}\hlstd{(}\hlkwc{average} \hlstd{= H} \hlopt{/} \hlstd{AB)} \hlopt{%>%}
    \hlkwd{filter}\hlstd{(salary} \hlopt{>} \hlnum{0}\hlstd{, AB} \hlopt{>=} \hlnum{50}\hlstd{,} \hlopt{!}\hlstd{(playerID} \hlopt{%in%} \hlstd{Pitching}\hlopt{$}\hlstd{playerID))}
\hlstd{merged}
\end{alltt}
\begin{verbatim}
## Source: local data frame [10,576 x 26]
## 
##     playerID yearID stint teamID lgID   G G_batting  AB  R   H X2B X3B HR
## 1  abbotje01   1998     1    CHA   AL  89        89 244 33  68  14   1 12
## 2  abbotje01   1999     1    CHA   AL  17        17  57  5   9   0   0  2
## 3  abbotje01   2000     1    CHA   AL  80        80 215 31  59  15   1  3
## 4  abbotku01   1993     1    OAK   AL  20        20  61 11  15   1   0  3
## 5  abbotku01   1994     1    FLO   NL 101       101 345 41  86  17   3  9
## 6  abbotku01   1995     1    FLO   NL 120       120 420 60 107  18   7 17
## 7  abbotku01   1996     1    FLO   NL 109       109 320 37  81  18   7  8
## 8  abbotku01   1997     1    FLO   NL  94        94 252 35  69  18   2  6
## 9  abbotku01   1998     1    OAK   AL  35        35 123 17  33   7   1  2
## 10 abbotku01   1999     1    COL   NL  96        96 286 41  78  17   2  8
## ..       ...    ...   ...    ...  ... ...       ... ... .. ... ... ... ..
## Variables not shown: RBI (int), SB (int), CS (int), BB (int), SO (int),
##   IBB (int), HBP (int), SH (int), SF (int), GIDP (int), G_old (int),
##   salary (int), average (dbl)
\end{verbatim}
\end{kframe}
\end{knitrout}

Here I've used \pkg{dplyr} to perform a few filtering and processing operations to set up the analysis. I merged Lahman's \code{Batting} dataset, which contains statistics on batting performance per player per season, with player salary information. I computed each player's batting average in each season: the fraction of times at bat ("AB") that led to a hit ("H"). I filtered for cases where the player had a nonzero salary, was at bat at least 50 times, and was not a pitcher (pitchers in baseball are typically not paid based on their batting ability).

Suppose we wish to investigate how players' salaries are related to their batting averages. We could start by performing a single linear regression predicting the log of salary based on batting average, including the year as a second predictor. A scatterplot shows that this relationship is plausible:

\begin{center}
\begin{knitrout}
\definecolor{shadecolor}{rgb}{0.969, 0.969, 0.969}\color{fgcolor}\begin{kframe}
\begin{alltt}
\hlkwd{library}\hlstd{(ggplot2)}
\hlkwd{ggplot}\hlstd{(merged,} \hlkwd{aes}\hlstd{(average, salary,} \hlkwc{color} \hlstd{= yearID))} \hlopt{+} \hlkwd{geom_point}\hlstd{()} \hlopt{+}
    \hlkwd{scale_y_log10}\hlstd{()}
\end{alltt}
\end{kframe}
\includegraphics[width=3in,height=3in]{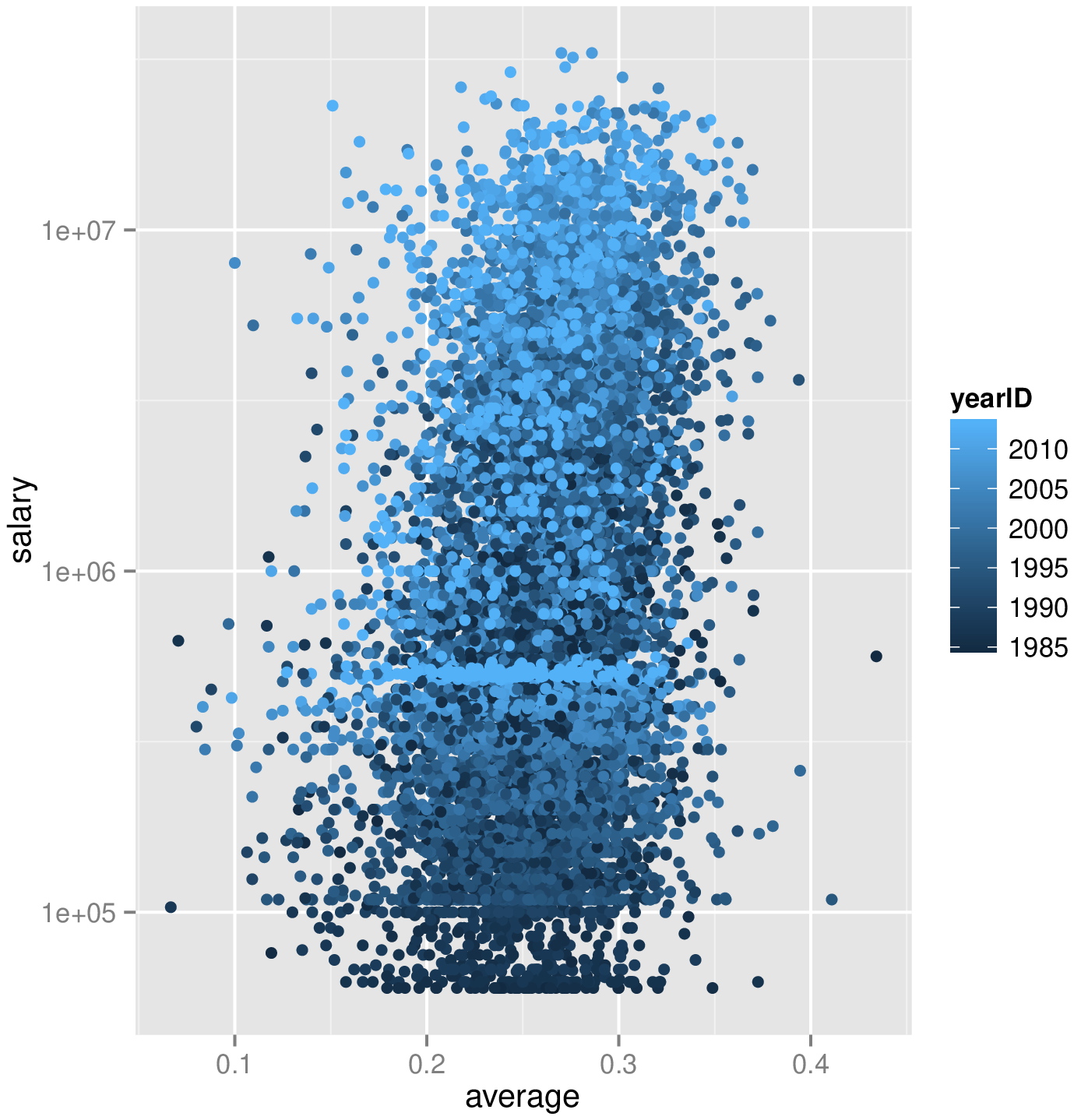} 

\end{knitrout}
\end{center}

\begin{knitrout}
\definecolor{shadecolor}{rgb}{0.969, 0.969, 0.969}\color{fgcolor}\begin{kframe}
\begin{alltt}
\hlstd{salary_fit} \hlkwb{<-} \hlkwd{lm}\hlstd{(}\hlkwd{log10}\hlstd{(salary)} \hlopt{~} \hlstd{average} \hlopt{+} \hlstd{yearID, merged)}
\hlkwd{summary}\hlstd{(salary_fit)}
\end{alltt}
\begin{verbatim}
## 
## Call:
## lm(formula = log10(salary) ~ average + yearID, data = merged)
## 
## Residuals:
##      Min       1Q   Median       3Q      Max 
## -1.32065 -0.48538 -0.01891  0.44176  1.44202 
## 
## Coefficients:
##               Estimate Std. Error t value Pr(>|t|)    
## (Intercept) -6.055e+01  1.255e+00  -48.23   <2e-16 ***
## average      3.743e+00  1.328e-01   28.18   <2e-16 ***
## yearID       3.277e-02  6.278e-04   52.19   <2e-16 ***
## ---
## Signif. codes:  0 '***' 0.001 '**' 0.01 '*' 0.05 '.' 0.1 ' ' 1
## 
## Residual standard error: 0.5252 on 10573 degrees of freedom
## Multiple R-squared:  0.2513,	Adjusted R-squared:  0.2512 
## F-statistic:  1775 on 2 and 10573 DF,  p-value: < 2.2e-16
\end{verbatim}
\end{kframe}
\end{knitrout}

This finds both a player's batting average and the year to be positively related to salary. What if we hypothesize that different teams place different amounts of emphasis on batting average for determining salary? We may be interested in performing the regression within each team. This can be done using \code{group\_by} and \code{do}.

\begin{knitrout}
\definecolor{shadecolor}{rgb}{0.969, 0.969, 0.969}\color{fgcolor}\begin{kframe}
\begin{alltt}
\hlstd{team_regressions} \hlkwb{<-} \hlstd{merged} \hlopt{%>%} \hlkwd{group_by}\hlstd{(teamID)} \hlopt{%>%}
    \hlkwd{do}\hlstd{(}\hlkwd{tidy}\hlstd{(}\hlkwd{lm}\hlstd{(}\hlkwd{log10}\hlstd{(salary)} \hlopt{~} \hlstd{average} \hlopt{+} \hlstd{yearID, .),} \hlkwc{conf.int} \hlstd{=} \hlnum{TRUE}\hlstd{))}
\hlstd{team_regressions}
\end{alltt}
\begin{verbatim}
## Source: local data frame [105 x 8]
## Groups: teamID
## 
##    teamID        term      estimate    std.error  statistic      p.value
## 1     ANA (Intercept) -113.85204105 46.013954092  -2.474294 1.493615e-02
## 2     ANA     average    2.97881540  1.382006026   2.155429 3.339096e-02
## 3     ANA      yearID    0.05951026  0.023018726   2.585298 1.108800e-02
## 4     ARI (Intercept)  -49.17939069 15.715043429  -3.129447 2.015982e-03
## 5     ARI     average    1.05909227  0.982962692   1.077449 2.825906e-01
## 6     ARI      yearID    0.02741524  0.007820337   3.505634 5.633804e-04
## 7     ATL (Intercept)  -44.42790245  6.730631459  -6.600852 1.411298e-10
## 8     ATL     average    3.04433694  0.697931557   4.361942 1.671522e-05
## 9     ATL      yearID    0.02482295  0.003374796   7.355394 1.225674e-12
## 10    BAL (Intercept)  -66.54304889  6.374462314 -10.439006 1.249081e-22
## ..    ...         ...           ...          ...        ...          ...
## Variables not shown: conf.low (dbl), conf.high (dbl)
\end{verbatim}
\end{kframe}
\end{knitrout}

With the regressions in this tidied format, we can examine and visualize the effect size estimates and confidence interval across all teams:

\begin{center}
\begin{knitrout}
\definecolor{shadecolor}{rgb}{0.969, 0.969, 0.969}\color{fgcolor}\begin{kframe}
\begin{alltt}
\hlstd{coefs} \hlkwb{<-} \hlstd{team_regressions} \hlopt{%>%} \hlkwd{ungroup}\hlstd{()} \hlopt{%>%}
    \hlkwd{filter}\hlstd{(term} \hlopt{==} \hlstr{"average"}\hlstd{)} \hlopt{%>%}
    \hlkwd{mutate}\hlstd{(}\hlkwc{teamID} \hlstd{=} \hlkwd{reorder}\hlstd{(teamID, estimate))}
\hlkwd{ggplot}\hlstd{(coefs,} \hlkwd{aes}\hlstd{(}\hlkwc{x} \hlstd{= estimate,} \hlkwc{y} \hlstd{= teamID))} \hlopt{+} \hlkwd{geom_point}\hlstd{()} \hlopt{+}
    \hlkwd{geom_errorbarh}\hlstd{(}\hlkwd{aes}\hlstd{(}\hlkwc{xmin} \hlstd{= conf.low,} \hlkwc{xmax} \hlstd{= conf.high))} \hlopt{+}
    \hlkwd{geom_vline}\hlstd{(}\hlkwc{color} \hlstd{=} \hlstr{"red"}\hlstd{)}
\end{alltt}
\end{kframe}
\includegraphics[width=3in,height=3in]{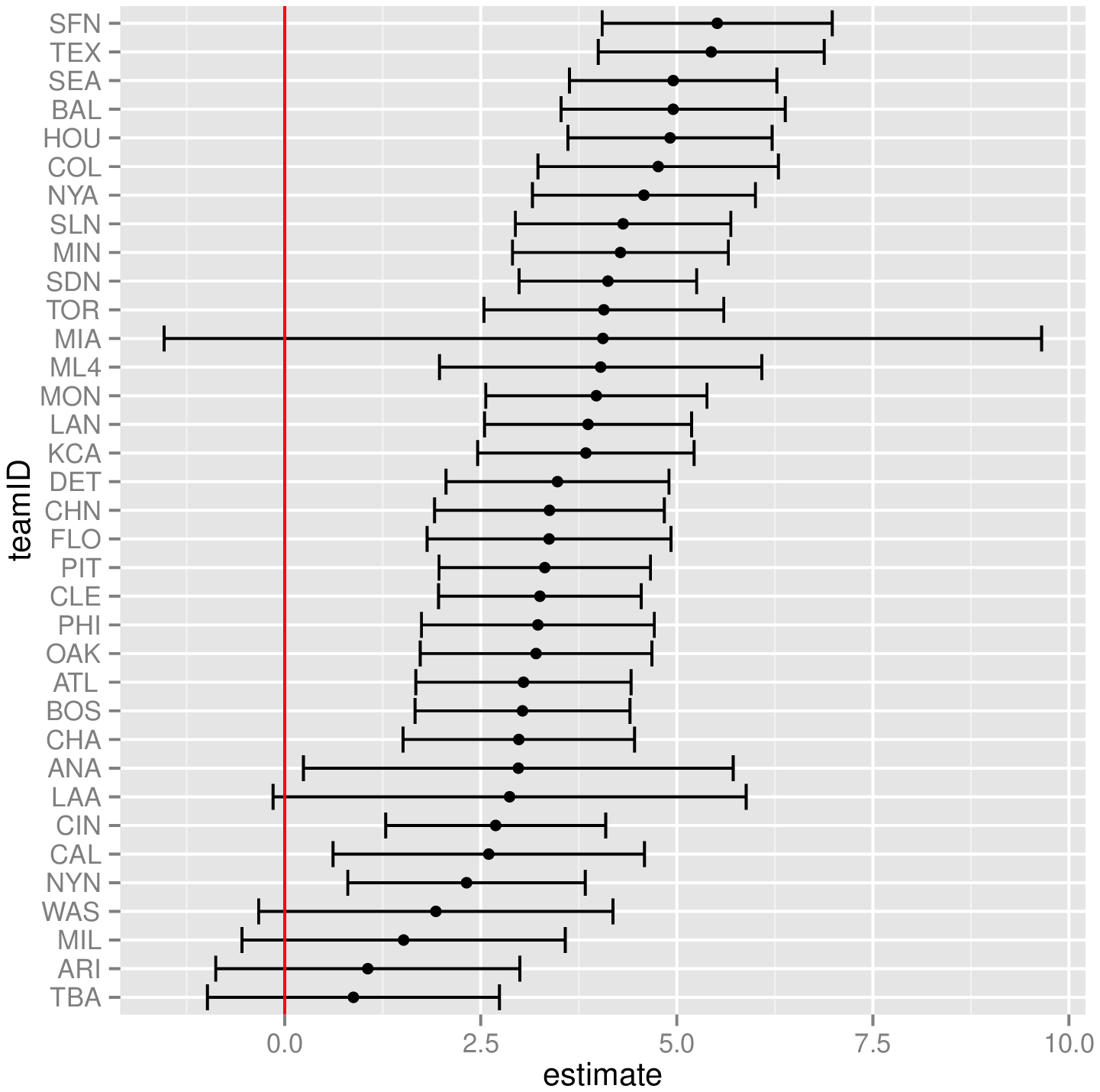} 

\end{knitrout}
\end{center}

Any analysis that has a tidying method can easily be performed within each team. For example, rather than performing a linear regression, we could find the Spearman correlation between salary and batting average within each team:

\begin{knitrout}
\definecolor{shadecolor}{rgb}{0.969, 0.969, 0.969}\color{fgcolor}\begin{kframe}
\begin{alltt}
\hlstd{cors} \hlkwb{<-} \hlstd{merged} \hlopt{%>%} \hlkwd{group_by}\hlstd{(teamID)} \hlopt{%>%}
    \hlkwd{do}\hlstd{(}\hlkwd{tidy}\hlstd{(}\hlkwd{cor.test}\hlstd{(.}\hlopt{$}\hlstd{salary, .}\hlopt{$}\hlstd{average,} \hlkwc{method} \hlstd{=} \hlstr{"spearman"}\hlstd{)))} \hlopt{%>%}
    \hlstd{ungroup} \hlopt{%>%} \hlkwd{arrange}\hlstd{(estimate)}
\hlkwd{tail}\hlstd{(cors,} \hlnum{3}\hlstd{)}
\end{alltt}
\begin{verbatim}
## Source: local data frame [3 x 4]
## 
##   teamID  estimate statistic      p.value
## 1    SFN 0.3455152   5615313 7.188991e-12
## 2    SLN 0.3480923   5196985 8.869388e-12
## 3    TEX 0.3641821   4821506 1.227396e-12
\end{verbatim}
\end{kframe}
\end{knitrout}

One could just as easily perform ANOVA, time series analyses, or nonlinear regressions within each subgroup. Similarly, one could perform analyses while grouping by league, by year, or by player, rather than by team. These analyses of subgroups are made possible by \pkg{broom}'s tidying methods, since they allow the model outputs to be recombined into a new tidy output.

\subsection{Bootstrapped confidence and prediction intervals}

\label{sec:bootstrapping}

Since tidy outputs can be recombined across replicates, it is also well suited to bootstrapping and permutation tests. Bootstrapping consists of randomly sampling observations from a dataset with replacement, then performing the same analysis individually on each bootstrapped replicate. The variation in the resulting value is then a reasonable approximation of the standard error of the estimate \citep{efron1994introduction}.

Suppose we wish to fit a nonlinear model to the weight/mileage relationship in the \code{mtcars} dataset, which comes built-in to R.

\begin{center}
\begin{knitrout}
\definecolor{shadecolor}{rgb}{0.969, 0.969, 0.969}\color{fgcolor}\begin{kframe}
\begin{alltt}
\hlkwd{ggplot}\hlstd{(mtcars,} \hlkwd{aes}\hlstd{(mpg, wt))} \hlopt{+} \hlkwd{geom_point}\hlstd{()}
\end{alltt}
\end{kframe}
\includegraphics[width=3in,height=3in]{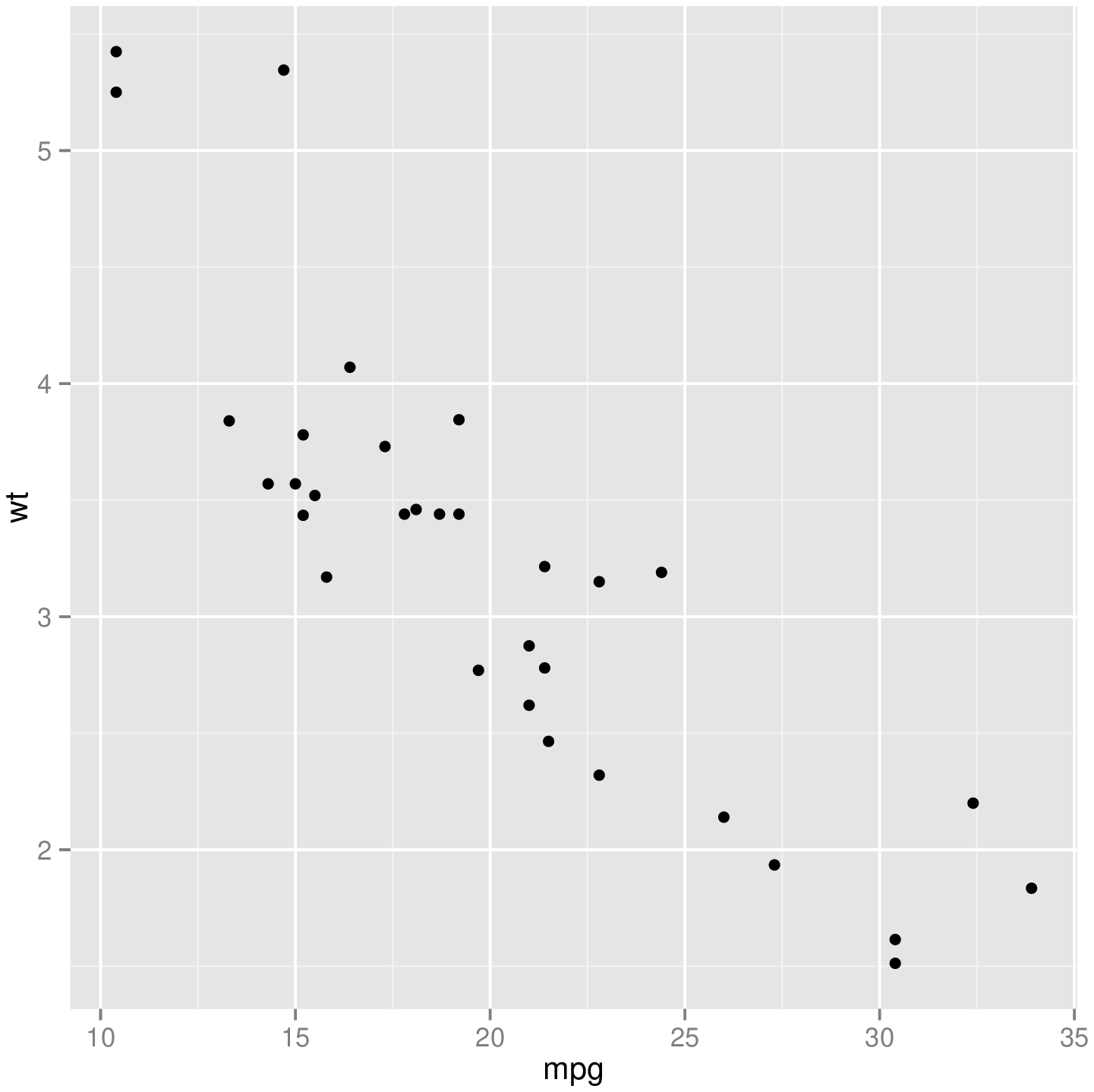} 

\end{knitrout}
\end{center}

We might use the method of nonlinear least squares (the built-in \code{nls} function) to fit a model.

\begin{center}
\begin{knitrout}
\definecolor{shadecolor}{rgb}{0.969, 0.969, 0.969}\color{fgcolor}\begin{kframe}
\begin{alltt}
\hlstd{nlsfit} \hlkwb{<-} \hlkwd{nls}\hlstd{(mpg} \hlopt{~} \hlstd{k} \hlopt{/} \hlstd{wt} \hlopt{+} \hlstd{b, mtcars,} \hlkwc{start}\hlstd{=}\hlkwd{list}\hlstd{(}\hlkwc{k}\hlstd{=}\hlnum{1}\hlstd{,} \hlkwc{b}\hlstd{=}\hlnum{0}\hlstd{))}
\hlkwd{summary}\hlstd{(nlsfit)}
\end{alltt}
\begin{verbatim}
## 
## Formula: mpg ~ k/wt + b
## 
## Parameters:
##   Estimate Std. Error t value Pr(>|t|)    
## k   45.829      4.249  10.786 7.64e-12 ***
## b    4.386      1.536   2.855  0.00774 ** 
## ---
## Signif. codes:  0 '***' 0.001 '**' 0.01 '*' 0.05 '.' 0.1 ' ' 1
## 
## Residual standard error: 2.774 on 30 degrees of freedom
## 
## Number of iterations to convergence: 1 
## Achieved convergence tolerance: 2.877e-08
\end{verbatim}
\begin{alltt}
\hlkwd{confint}\hlstd{(nlsfit)}
\end{alltt}
\begin{verbatim}
##        2.5%     97.5%
## k 37.151556 54.507419
## b  1.248471  7.524038
\end{verbatim}
\begin{alltt}
\hlkwd{ggplot}\hlstd{(mtcars,} \hlkwd{aes}\hlstd{(wt, mpg))} \hlopt{+} \hlkwd{geom_point}\hlstd{()} \hlopt{+} \hlkwd{geom_line}\hlstd{(}\hlkwd{aes}\hlstd{(}\hlkwc{y}\hlstd{=}\hlkwd{predict}\hlstd{(nlsfit)))}
\end{alltt}
\end{kframe}
\includegraphics[width=3in,height=3in]{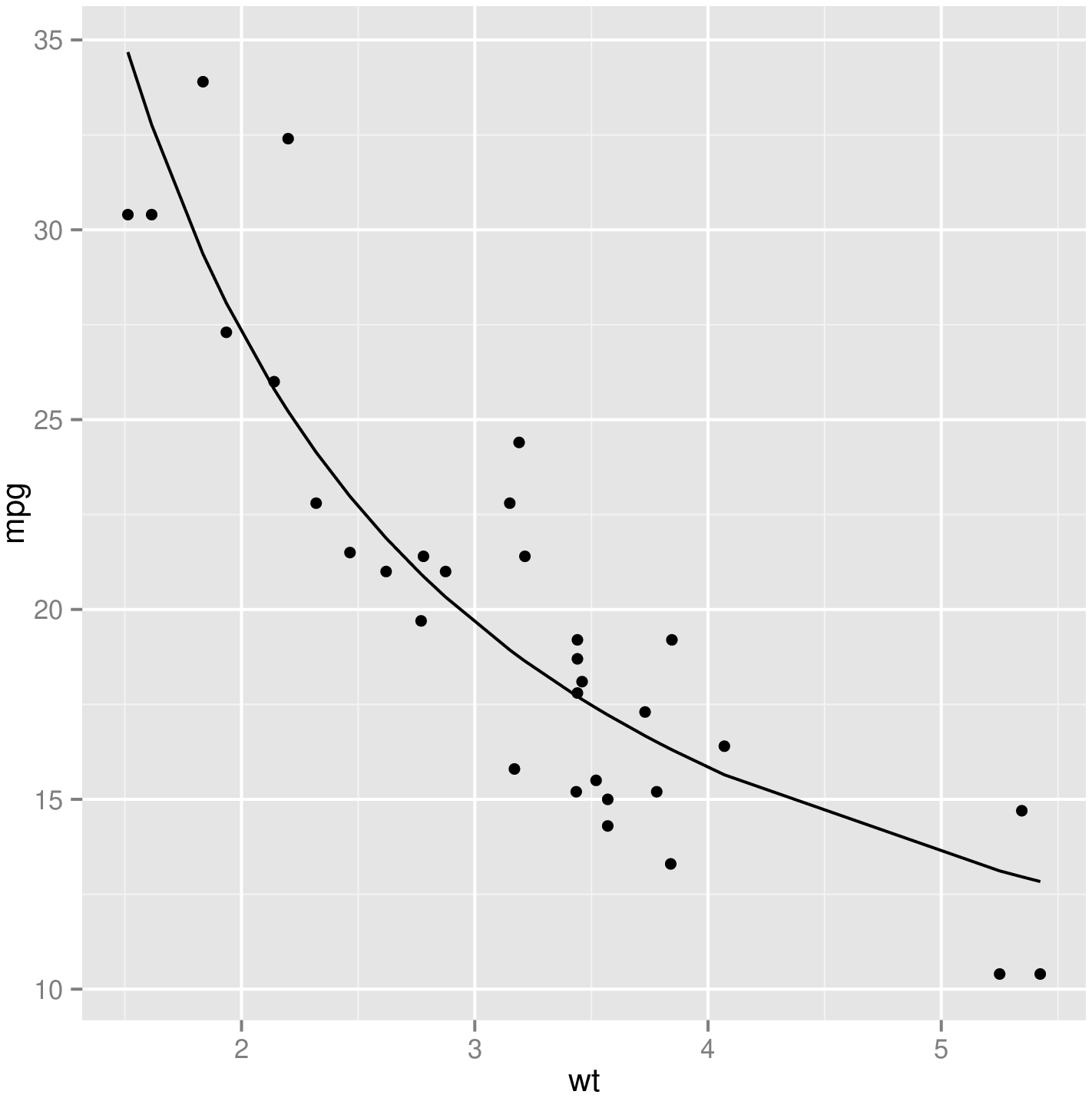} 

\end{knitrout}
\end{center}

While this does provide p-values and confidence intervals for the parameters, these are based on model assumptions that may not hold in real data. Bootstrapping is a popular method for providing confidence intervals and predictions that are more robust to the nature of the data.

The \textbf{broom} package provides a \code{bootstrap} function that in combination with \pkg{dplyr} and a tidying method makes bootstrapping straightforward. \code{bootstrap(.data, B)} wraps a data table so that the next \code{do} step occurs $B$ times, each time operating on the data resampled with replacement. Within each of these \code{do} applications, we construct a tidied version of a nonlinear least squares fit.

\begin{knitrout}
\definecolor{shadecolor}{rgb}{0.969, 0.969, 0.969}\color{fgcolor}\begin{kframe}
\begin{alltt}
\hlkwd{set.seed}\hlstd{(}\hlnum{2014}\hlstd{)}
\hlstd{bootnls} \hlkwb{<-} \hlstd{mtcars} \hlopt{%>%} \hlkwd{bootstrap}\hlstd{(}\hlnum{500}\hlstd{)} \hlopt{%>%}
    \hlkwd{do}\hlstd{(}\hlkwd{tidy}\hlstd{(}\hlkwd{nls}\hlstd{(mpg} \hlopt{~} \hlstd{k} \hlopt{/} \hlstd{wt} \hlopt{+} \hlstd{b, .,} \hlkwc{start}\hlstd{=}\hlkwd{list}\hlstd{(}\hlkwc{k}\hlstd{=}\hlnum{1}\hlstd{,} \hlkwc{b}\hlstd{=}\hlnum{0}\hlstd{))))}
\end{alltt}
\end{kframe}
\end{knitrout}

This produces a summary of the coefficients from each replication, combined into one data frame:

\begin{knitrout}
\definecolor{shadecolor}{rgb}{0.969, 0.969, 0.969}\color{fgcolor}\begin{kframe}
\begin{alltt}
\hlkwd{head}\hlstd{(bootnls)}
\end{alltt}
\begin{verbatim}
## Source: local data frame [6 x 6]
## Groups: replicate
## 
##   replicate term  estimate std.error  statistic      p.value
## 1         1    k 46.632502  4.026016 11.5827898 1.343891e-12
## 2         1    b  4.360637  1.538267  2.8347730 8.129583e-03
## 3         2    k 54.182476  4.964976 10.9129374 5.756829e-12
## 4         2    b  1.004868  1.897196  0.5296599 6.002460e-01
## 5         3    k 43.257212  3.564860 12.1343382 4.222953e-13
## 6         3    b  4.833510  1.297909  3.7240748 8.101899e-04
\end{verbatim}
\end{kframe}
\end{knitrout}

We can then calculate confidence intervals by considering the quantiles of the bootstrapped estimates. This is often referred to as the percentile method of bootstrapping, though it is not the only way to construct a confidence interval from bootstrap replicates.

\begin{knitrout}
\definecolor{shadecolor}{rgb}{0.969, 0.969, 0.969}\color{fgcolor}\begin{kframe}
\begin{alltt}
\hlstd{alpha} \hlkwb{=} \hlnum{.05}
\hlstd{bootnls} \hlopt{%>%} \hlkwd{group_by}\hlstd{(term)} \hlopt{%>%}
    \hlkwd{summarize}\hlstd{(}\hlkwc{conf.low} \hlstd{=} \hlkwd{quantile}\hlstd{(estimate, alpha} \hlopt{/} \hlnum{2}\hlstd{),}
              \hlkwc{conf.high} \hlstd{=} \hlkwd{quantile}\hlstd{(estimate,} \hlnum{1} \hlopt{-} \hlstd{alpha} \hlopt{/} \hlnum{2}\hlstd{))}
\end{alltt}
\begin{verbatim}
## Source: local data frame [2 x 3]
## 
##   term  conf.low conf.high
## 1    b  0.214338  6.952325
## 2    k 38.492700 59.024352
\end{verbatim}
\end{kframe}
\end{knitrout}

Finally, we can visualize the uncertainty in the actual curve using \code{augment} on each replication. We can then summarize the quantiles within each time point to produce bootstrap confidence intervals at each point.

\begin{center}
\begin{knitrout}
\definecolor{shadecolor}{rgb}{0.969, 0.969, 0.969}\color{fgcolor}\begin{kframe}
\begin{alltt}
\hlkwd{set.seed}\hlstd{(}\hlnum{2014}\hlstd{)}
\hlstd{bootnls} \hlkwb{<-} \hlstd{mtcars} \hlopt{%>%} \hlkwd{bootstrap}\hlstd{(}\hlnum{500}\hlstd{)} \hlopt{%>%}
    \hlkwd{do}\hlstd{(}\hlkwd{augment}\hlstd{(}\hlkwd{nls}\hlstd{(mpg} \hlopt{~} \hlstd{k} \hlopt{/} \hlstd{wt} \hlopt{+} \hlstd{b, .,} \hlkwc{start}\hlstd{=}\hlkwd{list}\hlstd{(}\hlkwc{k}\hlstd{=}\hlnum{1}\hlstd{,} \hlkwc{b}\hlstd{=}\hlnum{0}\hlstd{)), .))}
\hlstd{alpha} \hlkwb{=} \hlnum{.05}
\hlstd{bootnls_bytime} \hlkwb{<-} \hlstd{bootnls} \hlopt{%>%} \hlkwd{group_by}\hlstd{(wt)} \hlopt{%>%}
    \hlkwd{summarize}\hlstd{(}\hlkwc{conf.low} \hlstd{=} \hlkwd{quantile}\hlstd{(.fitted, alpha} \hlopt{/} \hlnum{2}\hlstd{),}
              \hlkwc{conf.high} \hlstd{=} \hlkwd{quantile}\hlstd{(.fitted,} \hlnum{1} \hlopt{-} \hlstd{alpha} \hlopt{/} \hlnum{2}\hlstd{),}
              \hlkwc{median} \hlstd{=} \hlkwd{median}\hlstd{(.fitted))}
\hlkwd{head}\hlstd{(bootnls_bytime)}
\end{alltt}
\begin{verbatim}
## Source: local data frame [6 x 4]
## 
##      wt conf.low conf.high   median
## 1 1.513 31.69528  36.84963 33.63675
## 2 1.615 30.53525  35.73523 32.38457
## 3 1.835 28.14591  32.87095 29.93906
## 4 1.935 26.35532  30.45919 28.14018
## 5 2.140 24.53684  27.83956 25.84737
## 6 2.200 24.44731  27.59607 25.75246
\end{verbatim}
\begin{alltt}
\hlkwd{ggplot}\hlstd{(mtcars,} \hlkwd{aes}\hlstd{(wt))} \hlopt{+} \hlkwd{geom_point}\hlstd{(}\hlkwd{aes}\hlstd{(}\hlkwc{y} \hlstd{= mpg))} \hlopt{+}
    \hlkwd{geom_line}\hlstd{(}\hlkwd{aes}\hlstd{(}\hlkwc{y} \hlstd{= .fitted),} \hlkwc{data} \hlstd{=} \hlkwd{augment}\hlstd{(nlsfit))} \hlopt{+}
    \hlkwd{geom_ribbon}\hlstd{(}\hlkwd{aes}\hlstd{(}\hlkwc{ymin} \hlstd{= conf.low,} \hlkwc{ymax} \hlstd{= conf.high),}
                \hlkwc{data} \hlstd{= bootnls_bytime,} \hlkwc{alpha} \hlstd{=} \hlnum{.1}\hlstd{)}
\end{alltt}
\end{kframe}
\includegraphics[width=3in,height=3in]{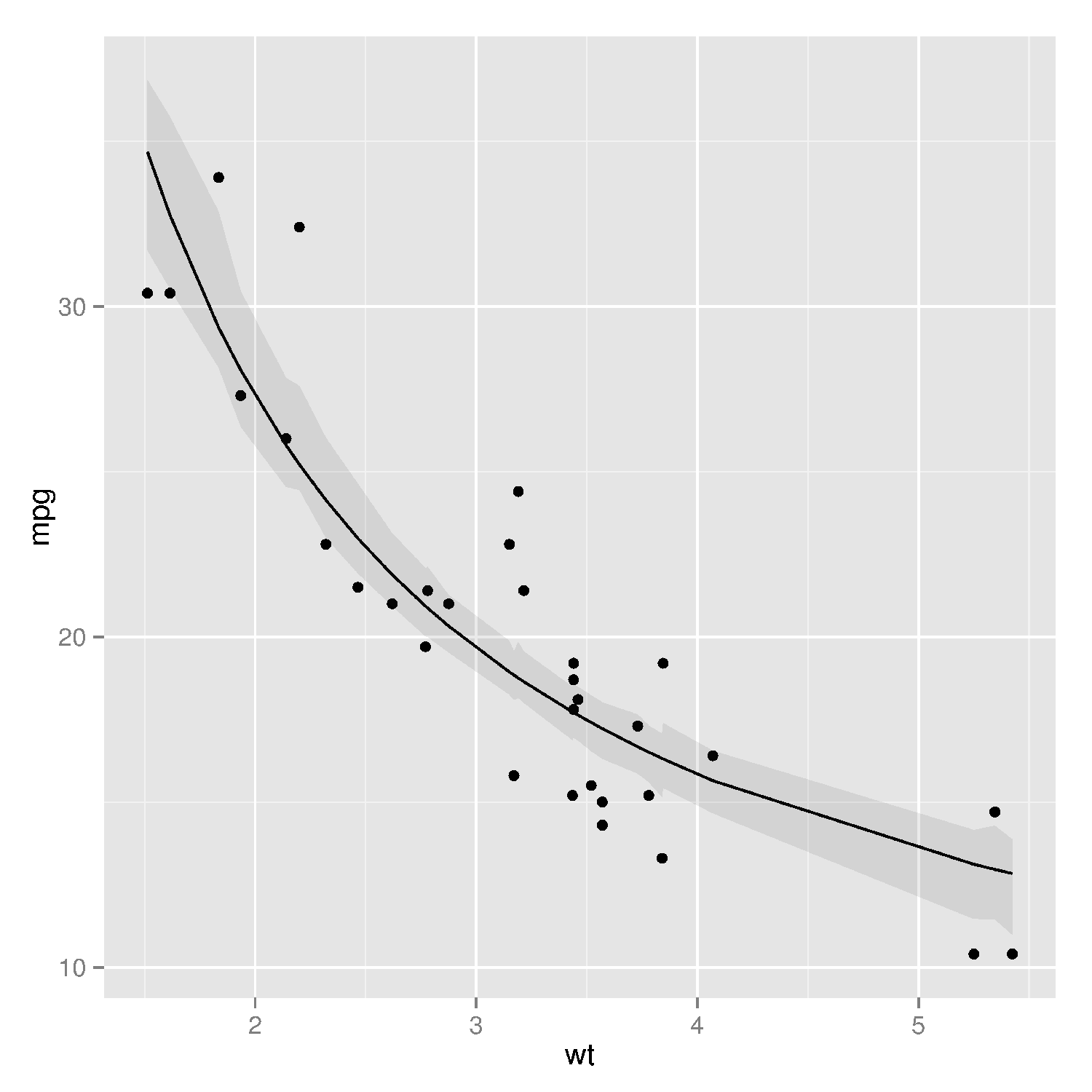} 

\end{knitrout}
\end{center}

This bootstrapping approach could be applied to any prediction function for which an \code{augment} method is defined. For example, we could use the built-in \pkg{splines} package to predict the curve using a natural cubic spline basis.

\begin{knitrout}
\definecolor{shadecolor}{rgb}{0.969, 0.969, 0.969}\color{fgcolor}\begin{kframe}
\begin{alltt}
\hlkwd{library}\hlstd{(splines)}
\hlstd{bootspline} \hlkwb{<-} \hlstd{mtcars} \hlopt{%>%} \hlkwd{bootstrap}\hlstd{(}\hlnum{500}\hlstd{)} \hlopt{%>%}
    \hlkwd{do}\hlstd{(}\hlkwd{augment}\hlstd{(}\hlkwd{lm}\hlstd{(mpg} \hlopt{~} \hlkwd{ns}\hlstd{(wt,} \hlnum{4}\hlstd{), .), .))}
\end{alltt}
\end{kframe}
\end{knitrout}

Since the bootstrap results are in the same format as the NLS fit, it is easy to combine them and then compare them on the same axis.

\begin{knitrout}
\definecolor{shadecolor}{rgb}{0.969, 0.969, 0.969}\color{fgcolor}\begin{kframe}
\begin{alltt}
\hlstd{bootnls}\hlopt{$}\hlstd{method} \hlkwb{<-} \hlstr{"nls"}
\hlstd{bootspline}\hlopt{$}\hlstd{method} \hlkwb{<-} \hlstr{"spline"}
\hlstd{allboot} \hlkwb{<-} \hlkwd{rbind_list}\hlstd{(bootnls, bootspline)}
\hlstd{allboot_bywt} \hlkwb{<-} \hlstd{allboot} \hlopt{%>%} \hlkwd{group_by}\hlstd{(wt, method)} \hlopt{%>%}
    \hlkwd{summarize}\hlstd{(}\hlkwc{conf.low} \hlstd{=} \hlkwd{quantile}\hlstd{(.fitted, alpha} \hlopt{/} \hlnum{2}\hlstd{),}
              \hlkwc{conf.high} \hlstd{=} \hlkwd{quantile}\hlstd{(.fitted,} \hlnum{1} \hlopt{-} \hlstd{alpha} \hlopt{/} \hlnum{2}\hlstd{),}
              \hlkwc{median} \hlstd{=} \hlkwd{median}\hlstd{(.fitted))}
\hlkwd{ggplot}\hlstd{(allboot_bywt,} \hlkwd{aes}\hlstd{(wt, median,} \hlkwc{color} \hlstd{= method))} \hlopt{+}
    \hlkwd{geom_line}\hlstd{()} \hlopt{+}
    \hlkwd{geom_ribbon}\hlstd{(}\hlkwd{aes}\hlstd{(}\hlkwc{ymin} \hlstd{= conf.low,} \hlkwc{ymax} \hlstd{= conf.high),} \hlkwc{lty} \hlstd{=} \hlnum{2}\hlstd{,} \hlkwc{alpha} \hlstd{=} \hlnum{.1}\hlstd{)}
\end{alltt}
\end{kframe}
\includegraphics[width=3in,height=3in]{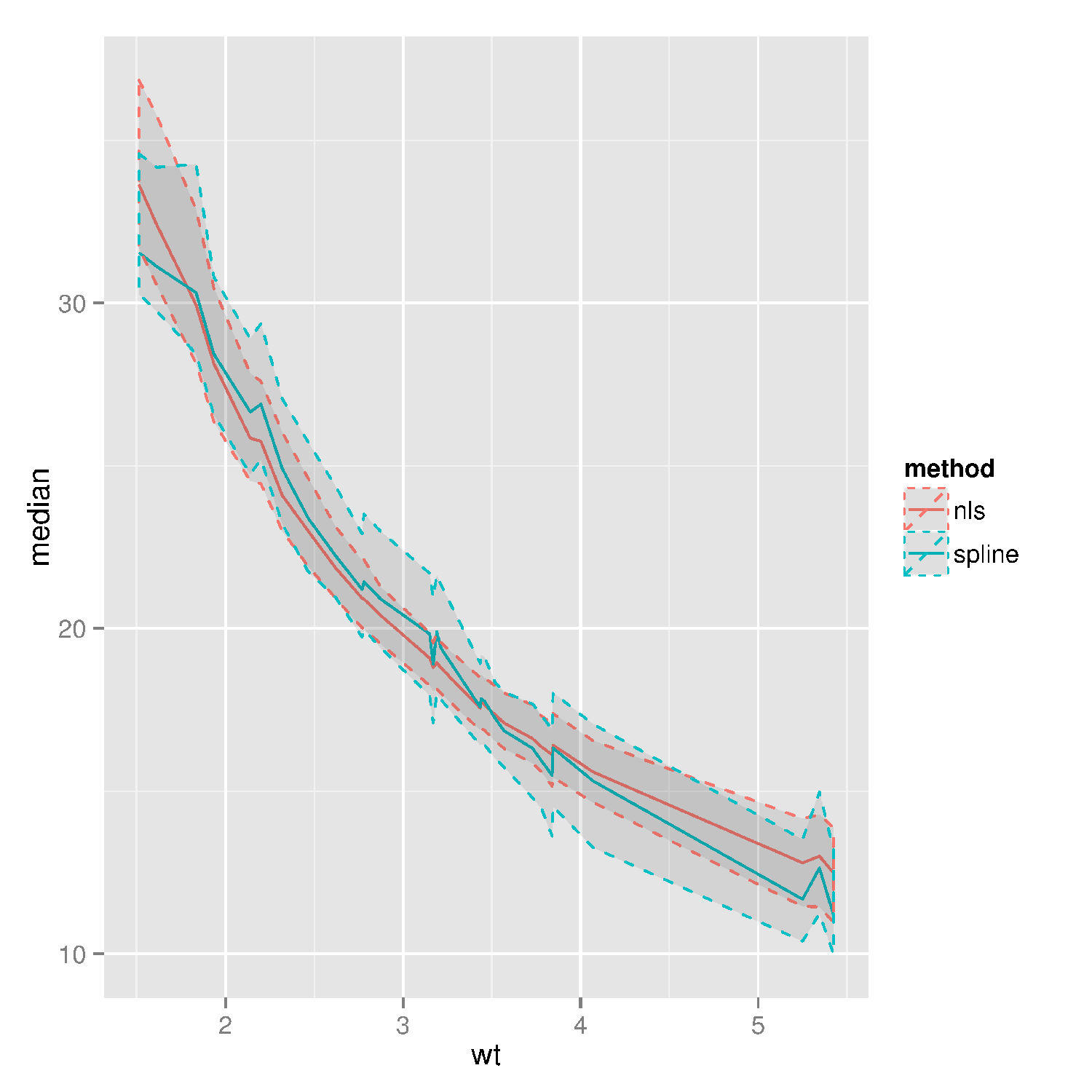} 

\end{knitrout}

Bootstrapping can be applied in this way to any object that \pkg{broom} can tidy, such as survival models from the \pkg{survival} package. Here we consider the \code{lung} dataset, which records the survival of patients with advanced lung cancer over time. We use the \code{coxph} function to fit a Cox proportional hazards regression model to model the likelihood of survival based on the demographic features of age and sex, and use \code{survfit} to construct a survival curve based on these hazards \citep{Therneau:2000tk}. Just like we did on the \code{mtcars} dataset, we perform this analysis within each of 500 bootstrap resamplings of the dataset. 

\begin{knitrout}
\definecolor{shadecolor}{rgb}{0.969, 0.969, 0.969}\color{fgcolor}\begin{kframe}
\begin{alltt}
\hlkwd{library}\hlstd{(survival)}
\hlstd{bootstraps_survival} \hlkwb{<-} \hlstd{lung} \hlopt{%>%} \hlkwd{bootstrap}\hlstd{(}\hlnum{500}\hlstd{)} \hlopt{%>%}
    \hlkwd{do}\hlstd{(}\hlkwd{tidy}\hlstd{(}\hlkwd{survfit}\hlstd{(}\hlkwd{coxph}\hlstd{(}\hlkwd{Surv}\hlstd{(time, status)} \hlopt{~} \hlstd{age} \hlopt{+} \hlstd{sex, .))))}
\hlkwd{head}\hlstd{(bootstraps_survival)}
\end{alltt}
\begin{verbatim}
## Source: local data frame [6 x 9]
## Groups: replicate
## 
##   replicate time n.risk n.event n.censor  estimate   std.error conf.high
## 1         1   11    228       2        0 0.9916307 0.005946493 1.0000000
## 2         1   12    226       1        0 0.9874253 0.007312389 1.0000000
## 3         1   13    225       3        0 0.9747625 0.010452461 0.9949378
## 4         1   15    222       2        0 0.9662883 0.012150010 0.9895752
## 5         1   26    220       1        0 0.9620413 0.012929476 0.9867322
## 6         1   30    219       1        0 0.9577853 0.013674363 0.9838023
## Variables not shown: conf.low (dbl)
\end{verbatim}
\end{kframe}
\end{knitrout}

We can then construct and visualize a confidence interval for the fraction of survival at each time point, just as we did for the NLS and spline fits.

\begin{center}
\begin{knitrout}
\definecolor{shadecolor}{rgb}{0.969, 0.969, 0.969}\color{fgcolor}\begin{kframe}
\begin{alltt}
\hlstd{alpha} \hlkwb{=} \hlnum{.05}
\hlstd{bootstraps_bytime} \hlkwb{<-} \hlstd{bootstraps_survival} \hlopt{%>%} \hlkwd{group_by}\hlstd{(time)} \hlopt{%>%}
    \hlkwd{summarize}\hlstd{(}\hlkwc{conf.low} \hlstd{=} \hlkwd{quantile}\hlstd{(estimate, alpha} \hlopt{/} \hlnum{2}\hlstd{),}
              \hlkwc{conf.high} \hlstd{=} \hlkwd{quantile}\hlstd{(estimate,} \hlnum{1} \hlopt{-} \hlstd{alpha} \hlopt{/} \hlnum{2}\hlstd{),}
              \hlkwc{estimate} \hlstd{=} \hlkwd{median}\hlstd{(estimate))}

\hlkwd{ggplot}\hlstd{(bootstraps_bytime,} \hlkwd{aes}\hlstd{(time, estimate))} \hlopt{+}
    \hlkwd{geom_line}\hlstd{()} \hlopt{+}
    \hlkwd{geom_ribbon}\hlstd{(}\hlkwd{aes}\hlstd{(}\hlkwc{ymin} \hlstd{= conf.low,} \hlkwc{ymax} \hlstd{= conf.high),} \hlkwc{lty} \hlstd{=} \hlnum{2}\hlstd{,} \hlkwc{alpha} \hlstd{=} \hlnum{.2}\hlstd{)}
\end{alltt}
\end{kframe}
\includegraphics[width=3in,height=3in]{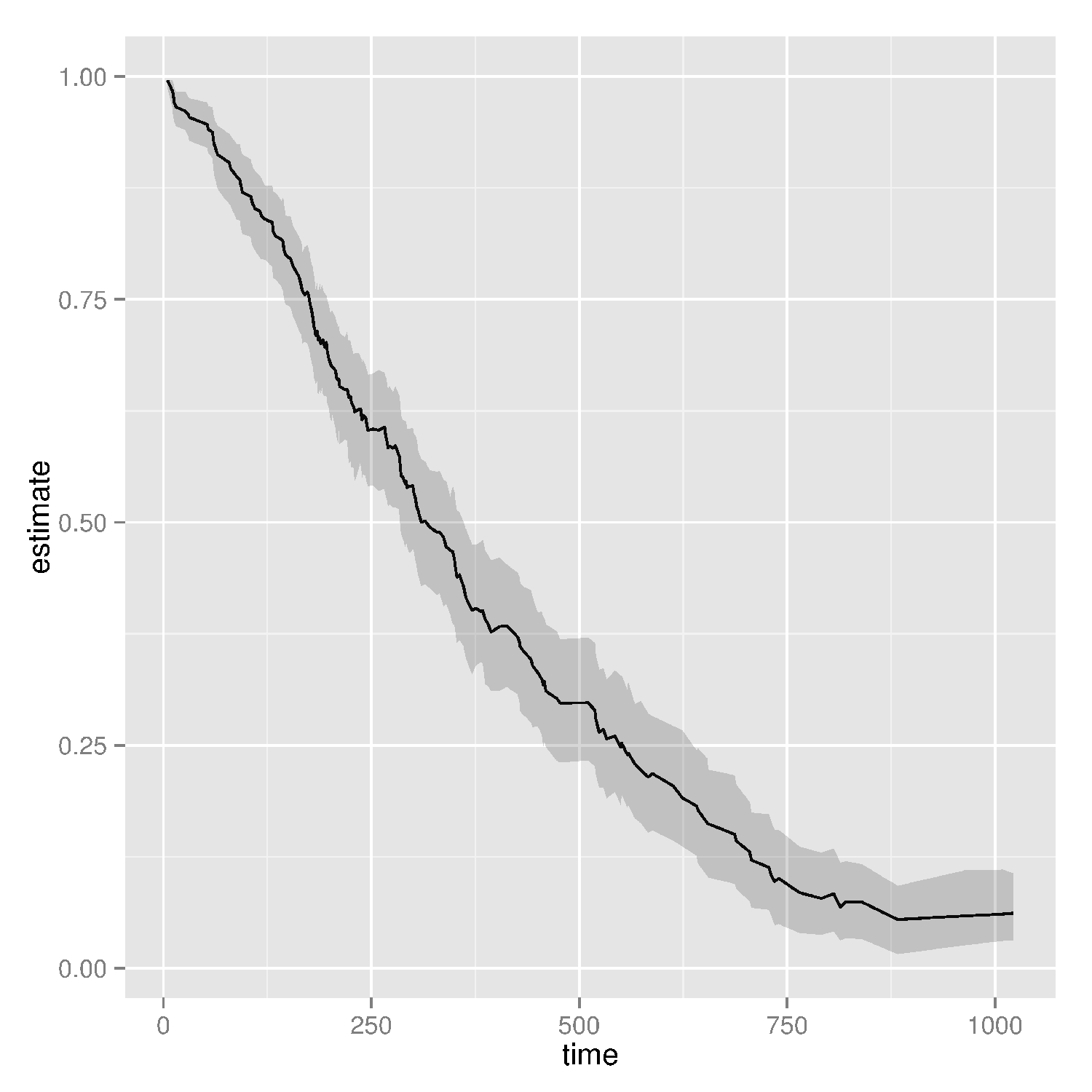} 

\end{knitrout}
\end{center}

It is worth noting that the tidy approach to bootstrapping may not necessarily be the most computationally efficient option for all modeling approaches. Some bootstrapping problems can be simplified to matrix operations that allow more very efficient processing and storage. Framing bootstrapping as a problem of recombining many tidy outputs is useful, however, for its simplicity and universality, as these same idioms can be applied to bootstrap any model with \code{tidy} or \code{augment} methods.

\subsection{Simulation of k-means clustering}

\label{sec:kmeans}

Because the \pkg{broom} package makes it possible to recombine many analyses, it is also well suited to simulation for the purpose of exploring a method's behavior and robustness. Tidied model outputs can easily be recombined across varying input parameters, or across simulation replications, and lend themselves to downstream analysis and visualization. In this simulation we examine a k-means clustering problem, and observe how the results are affected by the choice of the number of clusters and by the within-cluster variation in the data.

K-means clustering divides a set of points in an $n$-dimensional space into $k$ centers by assigning each point to the cluster with the nearest of $k$ designated centers. This is well suited for clustering problems where the data is assumed to be approximately multivariate Gaussian distributed around each cluster. The \pkg{stats} package provides an implementation of k-means clustering (based, by default, on the algorithm of \citet{HartiganWong:1979}), which requires the choice of $k$ to be made beforehand.

We start by exploring the effect of the choice of $k$ on the behavior of the clustering algorithm. We first generate random 2-dimensional data with three centers, around which points are distributed according to a standard multivariate Gaussian distribution:

\begin{knitrout}
\definecolor{shadecolor}{rgb}{0.969, 0.969, 0.969}\color{fgcolor}\begin{kframe}
\begin{alltt}
\hlkwd{set.seed}\hlstd{(}\hlnum{2014}\hlstd{)}
\hlstd{centers} \hlkwb{<-} \hlkwd{data.frame}\hlstd{(}\hlkwc{oracle} \hlstd{=} \hlkwd{factor}\hlstd{(}\hlnum{1}\hlopt{:}\hlnum{3}\hlstd{),}
                      \hlkwc{size} \hlstd{=} \hlkwd{c}\hlstd{(}\hlnum{100}\hlstd{,} \hlnum{150}\hlstd{,} \hlnum{50}\hlstd{),}
                      \hlkwc{x1} \hlstd{=} \hlkwd{c}\hlstd{(}\hlnum{5}\hlstd{,} \hlnum{0}\hlstd{,} \hlopt{-}\hlnum{3}\hlstd{),}
                      \hlkwc{x2} \hlstd{=} \hlkwd{c}\hlstd{(}\hlopt{-}\hlnum{1}\hlstd{,} \hlnum{1}\hlstd{,} \hlopt{-}\hlnum{2}\hlstd{))}
\hlstd{kdat} \hlkwb{<-} \hlstd{centers} \hlopt{%>%}
    \hlkwd{group_by}\hlstd{(oracle)} \hlopt{%>%}
    \hlkwd{do}\hlstd{(}\hlkwd{data.frame}\hlstd{(}\hlkwc{x1} \hlstd{=} \hlkwd{rnorm}\hlstd{(.}\hlopt{$}\hlstd{size[}\hlnum{1}\hlstd{], .}\hlopt{$}\hlstd{x1[}\hlnum{1}\hlstd{]),}
                  \hlkwc{x2} \hlstd{=} \hlkwd{rnorm}\hlstd{(.}\hlopt{$}\hlstd{size[}\hlnum{1}\hlstd{], .}\hlopt{$}\hlstd{x2[}\hlnum{1}\hlstd{])))}
\end{alltt}
\end{kframe}
\end{knitrout}

Now suppose we would like to perform k-means clustering on this data, while varying the number of clusters $k$. The \code{inflate} function, provided by \pkg{broom}, is useful for this purpose: it expands a dataset to repeat its original data once for each factorial combination of parameters.

\begin{knitrout}
\definecolor{shadecolor}{rgb}{0.969, 0.969, 0.969}\color{fgcolor}\begin{kframe}
\begin{alltt}
\hlstd{d} \hlkwb{<-} \hlkwd{data.frame}\hlstd{(}\hlkwc{a} \hlstd{=} \hlnum{1}\hlopt{:}\hlnum{3}\hlstd{,} \hlkwc{b} \hlstd{=} \hlnum{8}\hlopt{:}\hlnum{10}\hlstd{)}
\hlstd{d} \hlopt{%>%} \hlkwd{inflate}\hlstd{(}\hlkwc{x} \hlstd{=} \hlkwd{c}\hlstd{(}\hlstr{"apple"}\hlstd{,} \hlstr{"orange"}\hlstd{),} \hlkwc{y} \hlstd{=} \hlkwd{c}\hlstd{(}\hlstr{"car"}\hlstd{,} \hlstr{"boat"}\hlstd{))}
\end{alltt}
\begin{verbatim}
## Source: local data frame [12 x 4]
## Groups: x, y
## 
##         x    y a  b
## 1   apple boat 1  8
## 2   apple boat 2  9
## 3   apple boat 3 10
## 4   apple  car 1  8
## 5   apple  car 2  9
## 6   apple  car 3 10
## 7  orange boat 1  8
## 8  orange boat 2  9
## 9  orange boat 3 10
## 10 orange  car 1  8
## 11 orange  car 2  9
## 12 orange  car 3 10
\end{verbatim}
\end{kframe}
\end{knitrout}

In this case, we perform clustering on the same data (\code{kdat}) with each value of $k$ in \code{1:9}. Note that to reduce the role of randomness in the clustering process, we set \code{nstart = 5} to the \code{kmeans} function.

\begin{knitrout}
\definecolor{shadecolor}{rgb}{0.969, 0.969, 0.969}\color{fgcolor}\begin{kframe}
\begin{alltt}
\hlstd{kclusts} \hlkwb{<-} \hlstd{kdat} \hlopt{%>%} \hlkwd{inflate}\hlstd{(}\hlkwc{k} \hlstd{=} \hlnum{1}\hlopt{:}\hlnum{9}\hlstd{)} \hlopt{%>%} \hlkwd{group_by}\hlstd{(k)} \hlopt{%>%}
    \hlkwd{do}\hlstd{(}\hlkwc{clust} \hlstd{=} \hlkwd{kmeans}\hlstd{(}\hlkwd{select}\hlstd{(., x1, x2), .}\hlopt{$}\hlstd{k[}\hlnum{1}\hlstd{],} \hlkwc{nstart} \hlstd{=} \hlnum{5}\hlstd{))}
\hlstd{kclusts}
\end{alltt}
\begin{verbatim}
## Source: local data frame [9 x 2]
## Groups: <by row>
## 
##   k       clust
## 1 1 <S3:kmeans>
## 2 2 <S3:kmeans>
## 3 3 <S3:kmeans>
## 4 4 <S3:kmeans>
## 5 5 <S3:kmeans>
## 6 6 <S3:kmeans>
## 7 7 <S3:kmeans>
## 8 8 <S3:kmeans>
## 9 9 <S3:kmeans>
\end{verbatim}
\end{kframe}
\end{knitrout}

There are three levels at which we can examine a k-means clustering. As is true of regressions and other statistical models, each of these levels describes a separate observational unit, and therefore is extracted by a different \pkg{broom} generic.

\begin{itemize}
\item \textbf{Component level}: The centers, size, and within sum-of-squares for each cluster; computed by \code{tidy}
\item \textbf{Observation level}: The assignment of each point to a cluster; computed by \code{augment}
\item \textbf{Model level}: The total within sum-of-squares and between sum-of-squares; computed by \code{glance}
\end{itemize}

We extract tidied versions of these three levels from each of the 9 clustering objects produced in the simulation, then recombine each level into a single large table containing the results for all values of $k$. We use the shortcut for performing tidying functions on a rowwise table first demonstrated in \nameref{sec:split_apply_combine}.

\begin{knitrout}
\definecolor{shadecolor}{rgb}{0.969, 0.969, 0.969}\color{fgcolor}\begin{kframe}
\begin{alltt}
\hlstd{clusters} \hlkwb{<-} \hlstd{kclusts} \hlopt{%>%} \hlkwd{tidy}\hlstd{(clust)}
\hlstd{assignments} \hlkwb{<-} \hlstd{kclusts} \hlopt{%>%} \hlkwd{augment}\hlstd{(clust, kdat)}
\hlstd{clusterings} \hlkwb{<-} \hlstd{kclusts} \hlopt{%>%} \hlkwd{glance}\hlstd{(clust)}
\end{alltt}
\end{kframe}
\end{knitrout}

The \code{assignments} object, generated using the \code{augment} method on each clustering, combines the cluster assignments across all values of $k$, thus allowing for simple visualization using faceting.

\begin{center}
\begin{knitrout}
\definecolor{shadecolor}{rgb}{0.969, 0.969, 0.969}\color{fgcolor}\begin{kframe}
\begin{alltt}
\hlkwd{head}\hlstd{(assignments)}
\end{alltt}
\begin{verbatim}
## Source: local data frame [6 x 5]
## Groups: k
## 
##   k oracle       x1         x2 .cluster
## 1 1      1 4.434320  0.5416470        1
## 2 1      1 5.321046 -0.9412882        1
## 3 1      1 5.125271 -1.5802282        1
## 4 1      1 6.353225 -1.6040549        1
## 5 1      1 3.712270 -3.4079344        1
## 6 1      1 5.322555 -0.7716317        1
\end{verbatim}
\begin{alltt}
\hlstd{p1} \hlkwb{<-} \hlkwd{ggplot}\hlstd{(assignments,} \hlkwd{aes}\hlstd{(x1, x2))} \hlopt{+}
    \hlkwd{geom_point}\hlstd{(}\hlkwd{aes}\hlstd{(}\hlkwc{color} \hlstd{= .cluster,} \hlkwc{shape} \hlstd{= oracle))} \hlopt{+}
    \hlkwd{facet_wrap}\hlstd{(}\hlopt{~} \hlstd{k)}
\hlstd{p1}
\end{alltt}
\end{kframe}
\includegraphics[width=3in,height=3in]{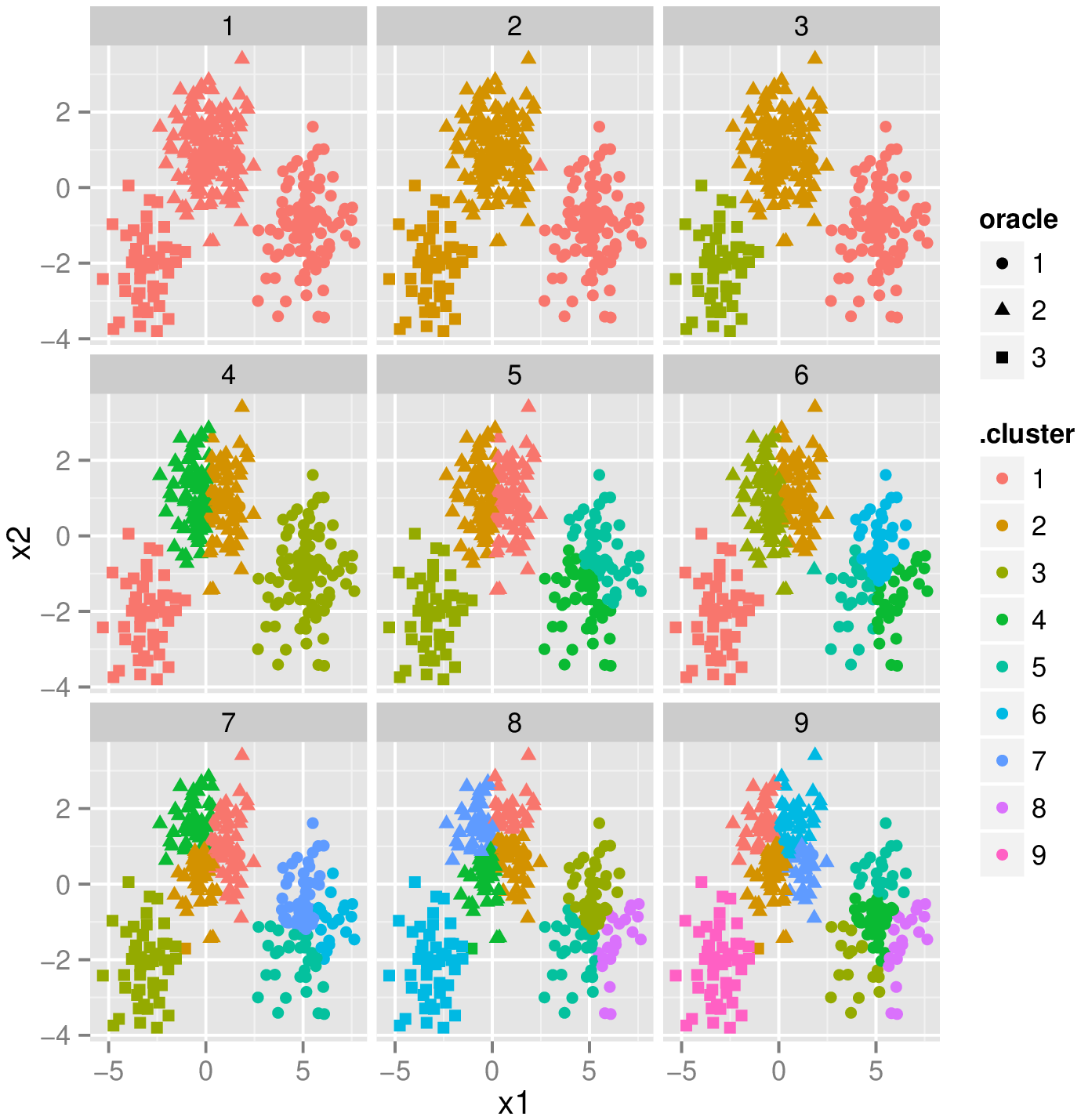} 

\end{knitrout}
\end{center}

We can use the results from the \code{tidy} outputs, which provide per-cluster information, to add the estimated center of each cluster to the plot.

\begin{center}
\begin{knitrout}
\definecolor{shadecolor}{rgb}{0.969, 0.969, 0.969}\color{fgcolor}\begin{kframe}
\begin{alltt}
\hlstd{p2} \hlkwb{<-} \hlstd{p1} \hlopt{+} \hlkwd{geom_point}\hlstd{(}\hlkwc{data} \hlstd{= clusters,} \hlkwc{size} \hlstd{=} \hlnum{10}\hlstd{,} \hlkwc{shape} \hlstd{=} \hlstr{"x"}\hlstd{)}
\hlstd{p2}
\end{alltt}
\end{kframe}
\includegraphics[width=3in,height=3in]{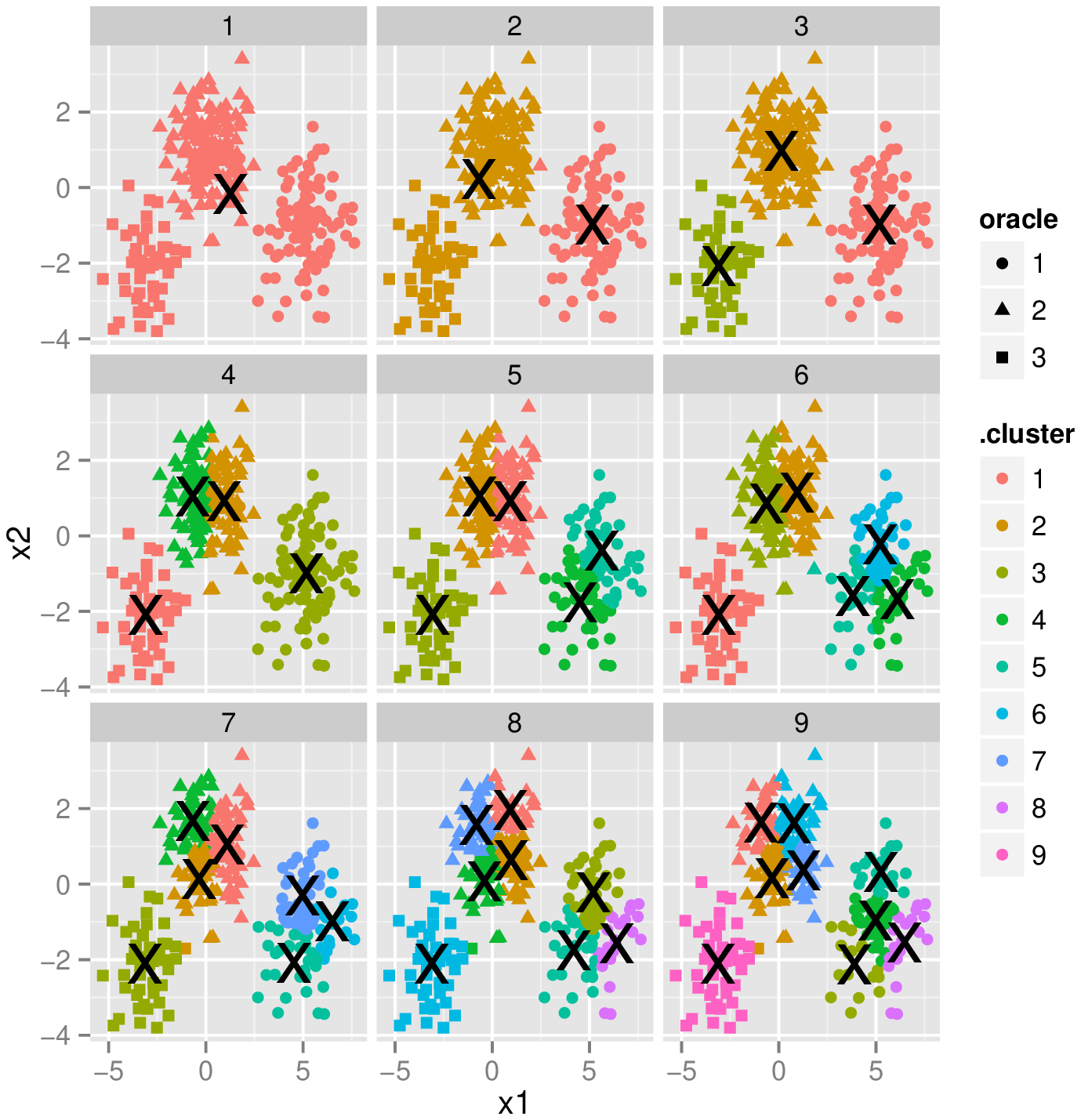} 

\end{knitrout}
\end{center}

Finally, we can examine the per-model data using the \code{glance} outputs. Of particular interest is the total within cluster sum-of-squares. The decrease in this within-cluster variation as clusters are added to the model can serve as a criterion for choosing $k$.

\begin{center}
\begin{knitrout}
\definecolor{shadecolor}{rgb}{0.969, 0.969, 0.969}\color{fgcolor}\begin{kframe}
\begin{alltt}
\hlkwd{ggplot}\hlstd{(clusterings,} \hlkwd{aes}\hlstd{(k, tot.withinss))} \hlopt{+} \hlkwd{geom_line}\hlstd{()}
\end{alltt}
\end{kframe}
\includegraphics[width=3in,height=3in]{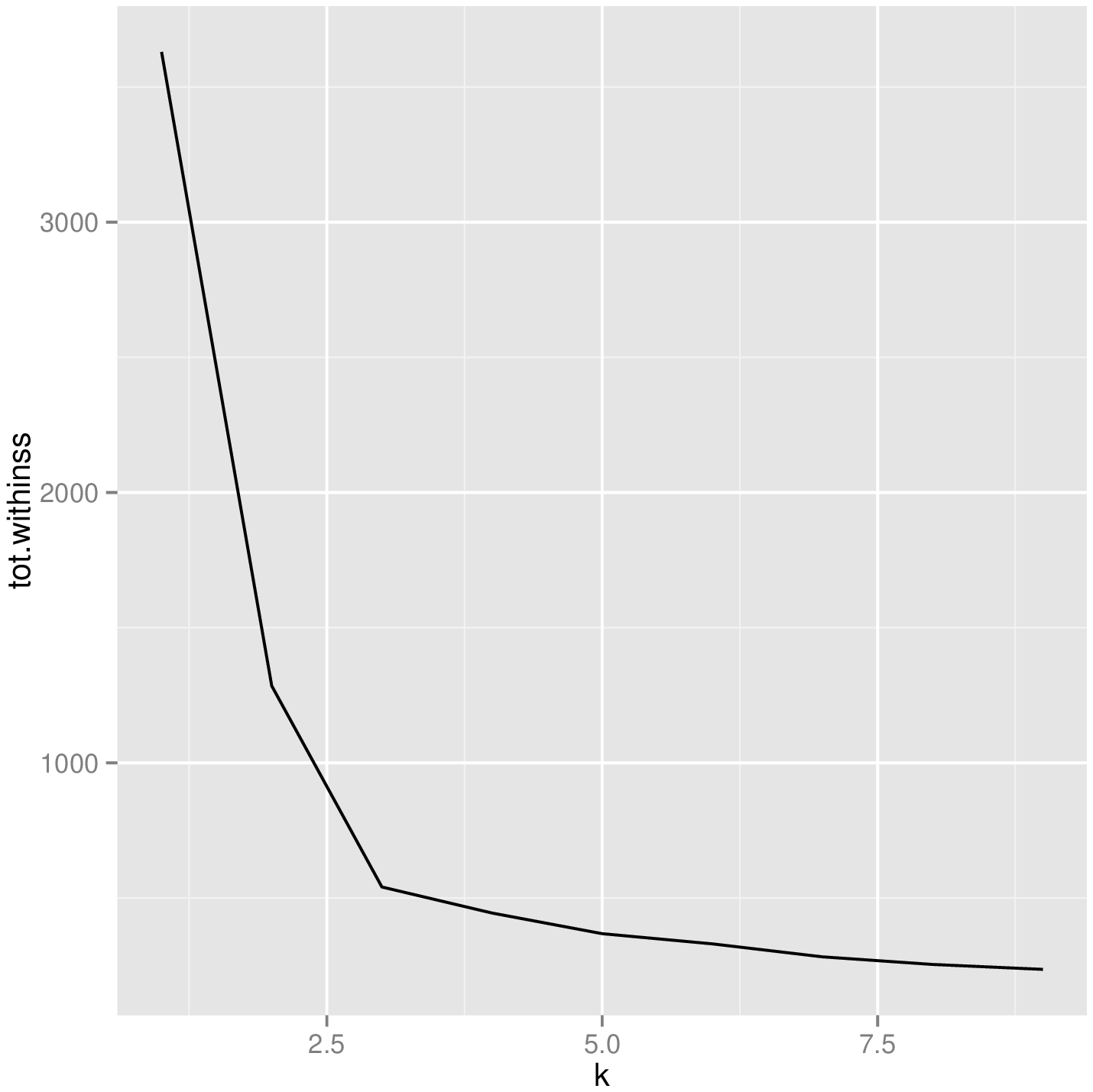} 

\end{knitrout}
\end{center}

The variance within each estimated cluster decreases as k increases, but one can notice a bend (or “elbow”) around the correct value of $k=3$. This bend indicates that additional clusters beyond the third have little value in terms of explaining the variation in the data. \citet{Tibshirani:2002fj} provides a more mathematically rigorous interpretation and implementation of a method to choose $k$ based on this profile. Note that all three methods of tidying data provided by \pkg{broom} play a separate role in visualization and analysis of these simulations.

While a simulation varying $k$ is useful, we may also be interested in how robust the algorithm is when the original data is altered. For example, the points around each center are generated from a multivariate normal distribution with standard deviation $\sigma=1$. If this standard deviation increases, making the cloud of points around each center more disperse, we would expect k-means clustering to be less accurate. We also wish to know the role of random variation, and therefore will perform 50 replicates of each simulation. These goals can be achieved with some small modifications to the previous simulation. We first generate the data using the same centers, but with multiple values of $\sigma$ and multiple independent replications:

\begin{knitrout}
\definecolor{shadecolor}{rgb}{0.969, 0.969, 0.969}\color{fgcolor}\begin{kframe}
\begin{alltt}
\hlkwd{set.seed}\hlstd{(}\hlnum{2014}\hlstd{)}
\hlstd{kdat_sd} \hlkwb{<-} \hlstd{centers} \hlopt{%>%}
    \hlkwd{inflate}\hlstd{(}\hlkwc{sd} \hlstd{=} \hlkwd{c}\hlstd{(}\hlnum{.5}\hlstd{,} \hlnum{1}\hlstd{,} \hlnum{2}\hlstd{,} \hlnum{4}\hlstd{),} \hlkwc{replication} \hlstd{=} \hlnum{1}\hlopt{:}\hlnum{50}\hlstd{)} \hlopt{%>%}
    \hlkwd{group_by}\hlstd{(oracle, sd, replication)} \hlopt{%>%}
    \hlkwd{do}\hlstd{(}\hlkwd{data.frame}\hlstd{(}\hlkwc{x1} \hlstd{=} \hlkwd{rnorm}\hlstd{(.}\hlopt{$}\hlstd{size[}\hlnum{1}\hlstd{], .}\hlopt{$}\hlstd{x1[}\hlnum{1}\hlstd{], .}\hlopt{$}\hlstd{sd[}\hlnum{1}\hlstd{]),}
                  \hlkwc{x2} \hlstd{=} \hlkwd{rnorm}\hlstd{(.}\hlopt{$}\hlstd{size[}\hlnum{1}\hlstd{], .}\hlopt{$}\hlstd{x2[}\hlnum{1}\hlstd{], .}\hlopt{$}\hlstd{sd[}\hlnum{1}\hlstd{])))}
\end{alltt}
\end{kframe}
\end{knitrout}

We perform k-means clustering 9 times on each dataset, choosing a different value of $k$ each time. We then extract the tidied, augmented, and glanced forms, which in this case are combined across all factorial combinations of \code{k}, \code{sd}, and \code{replication}. One could easily extend the simulation to alter other parameters such as the number and distribution of true cluster centers, or the \code{nstart} parameter.

\begin{knitrout}
\definecolor{shadecolor}{rgb}{0.969, 0.969, 0.969}\color{fgcolor}\begin{kframe}
\begin{alltt}
\hlstd{kclusts_sd} \hlkwb{<-} \hlstd{kdat_sd} \hlopt{%>%} \hlkwd{inflate}\hlstd{(}\hlkwc{k} \hlstd{=} \hlnum{1}\hlopt{:}\hlnum{9}\hlstd{)} \hlopt{%>%} \hlkwd{group_by}\hlstd{(k, sd, replication)} \hlopt{%>%}
    \hlkwd{do}\hlstd{(}\hlkwc{dat} \hlstd{= (.),} \hlkwc{clust} \hlstd{=} \hlkwd{kmeans}\hlstd{(}\hlkwd{select}\hlstd{(., x1, x2), .}\hlopt{$}\hlstd{k[}\hlnum{1}\hlstd{],} \hlkwc{nstart} \hlstd{=} \hlnum{5}\hlstd{))}
\hlstd{clusters_sd} \hlkwb{<-} \hlstd{kclusts_sd} \hlopt{%>%} \hlkwd{tidy}\hlstd{(clust)}
\hlstd{glances_sd} \hlkwb{<-} \hlstd{kclusts_sd} \hlopt{%>%} \hlkwd{glance}\hlstd{(clust)}
\hlstd{assignments_sd} \hlkwb{<-} \hlstd{kclusts_sd} \hlopt{%>%} \hlkwd{group_by}\hlstd{(k, sd, replication)} \hlopt{%>%}
    \hlkwd{do}\hlstd{(}\hlkwd{augment}\hlstd{(.}\hlopt{$}\hlstd{clust[[}\hlnum{1}\hlstd{]], .}\hlopt{$}\hlstd{dat[[}\hlnum{1}\hlstd{]]))}
\end{alltt}
\end{kframe}
\end{knitrout}

(Note that since the \code{augment} operation requires the original data, which changes in each replication, we needed a different approach to perform \code{augment} on each row). One interesting question is whether the cluster centers were estimated accurately in each simulation. The estimated centers are included in the recombined \code{tidy} output, which can be visualized alongside the true centers (red X's) separately for each value of $\sigma$:

\begin{center}
\begin{knitrout}
\definecolor{shadecolor}{rgb}{0.969, 0.969, 0.969}\color{fgcolor}\begin{kframe}
\begin{alltt}
\hlstd{clusters_sd} \hlopt{%>%} \hlkwd{filter}\hlstd{(k} \hlopt{==} \hlnum{3}\hlstd{)} \hlopt{%>%}
    \hlkwd{ggplot}\hlstd{(}\hlkwd{aes}\hlstd{(x1, x2))} \hlopt{+} \hlkwd{geom_point}\hlstd{()} \hlopt{+}
    \hlkwd{geom_point}\hlstd{(}\hlkwc{data} \hlstd{= centers,} \hlkwc{size} \hlstd{=} \hlnum{7}\hlstd{,} \hlkwc{color} \hlstd{=} \hlstr{"red"}\hlstd{,} \hlkwc{shape} \hlstd{=} \hlstr{"x"}\hlstd{)} \hlopt{+}
    \hlkwd{facet_wrap}\hlstd{(}\hlopt{~} \hlstd{sd)}
\end{alltt}
\end{kframe}
\includegraphics[width=3in,height=3in]{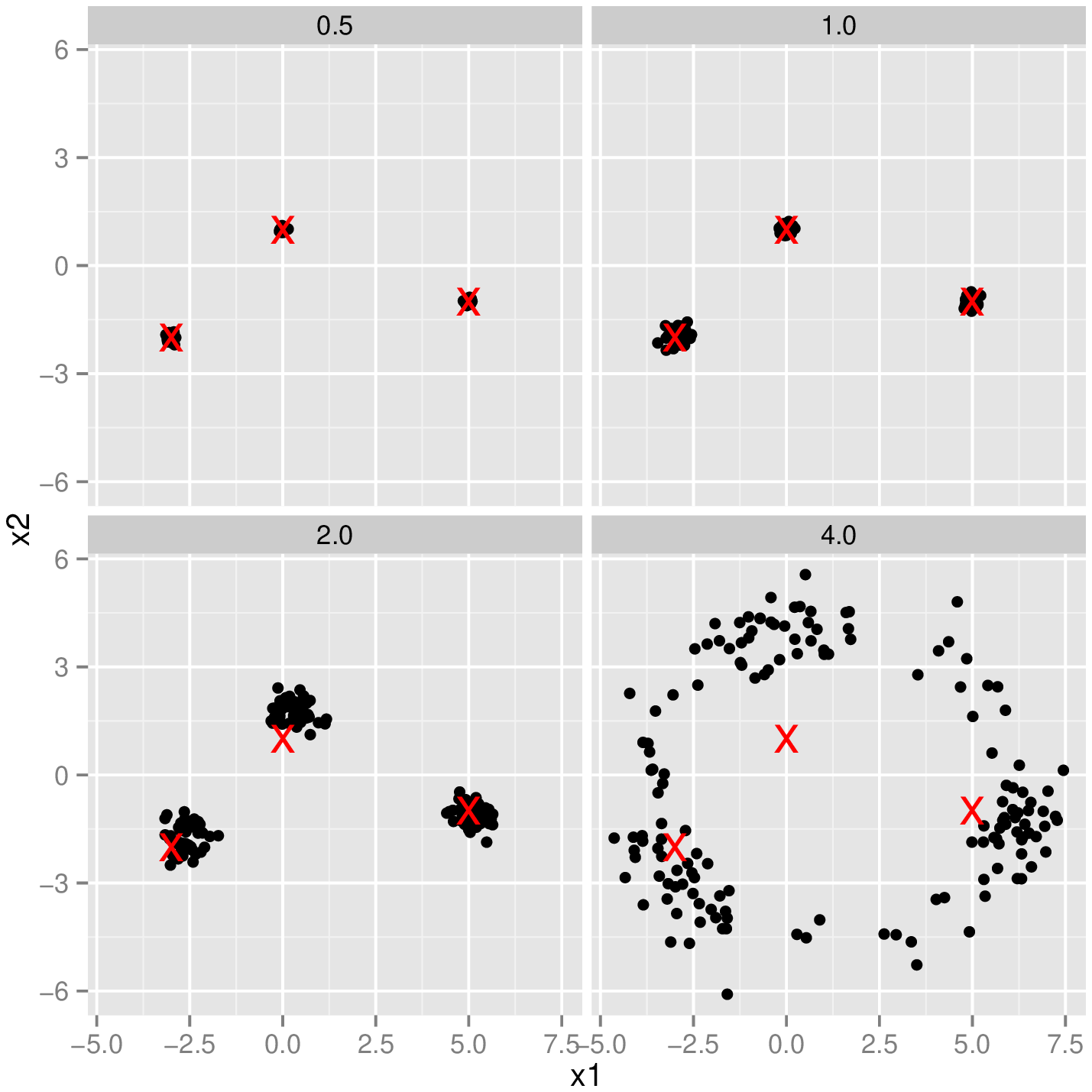} 

\end{knitrout}
\end{center}

We can see that the centers are estimated very accurately for $\sigma = .5$ and $1$, less accurately for $\sigma = 2$, and rather inaccurately for $\sigma = 4$. Also notably, the center estimates for $\sigma = 4$ are systematically biased  "outward" for two of the three clusters, as an artifact of the greater variation.

We can also see how the total within-sum-of-squares graph appears across all replications, and how it varies when changing the value of $\sigma$.

\begin{center}
\begin{knitrout}
\definecolor{shadecolor}{rgb}{0.969, 0.969, 0.969}\color{fgcolor}\begin{kframe}
\begin{alltt}
\hlkwd{ggplot}\hlstd{(glances_sd,} \hlkwd{aes}\hlstd{(k, tot.withinss,} \hlkwc{group} \hlstd{= replication))} \hlopt{+}
    \hlkwd{geom_line}\hlstd{()} \hlopt{+} \hlkwd{facet_wrap}\hlstd{(}\hlopt{~} \hlstd{sd,} \hlkwc{scales} \hlstd{=} \hlstr{"free_y"}\hlstd{)}
\end{alltt}
\end{kframe}
\includegraphics[width=3in,height=3in]{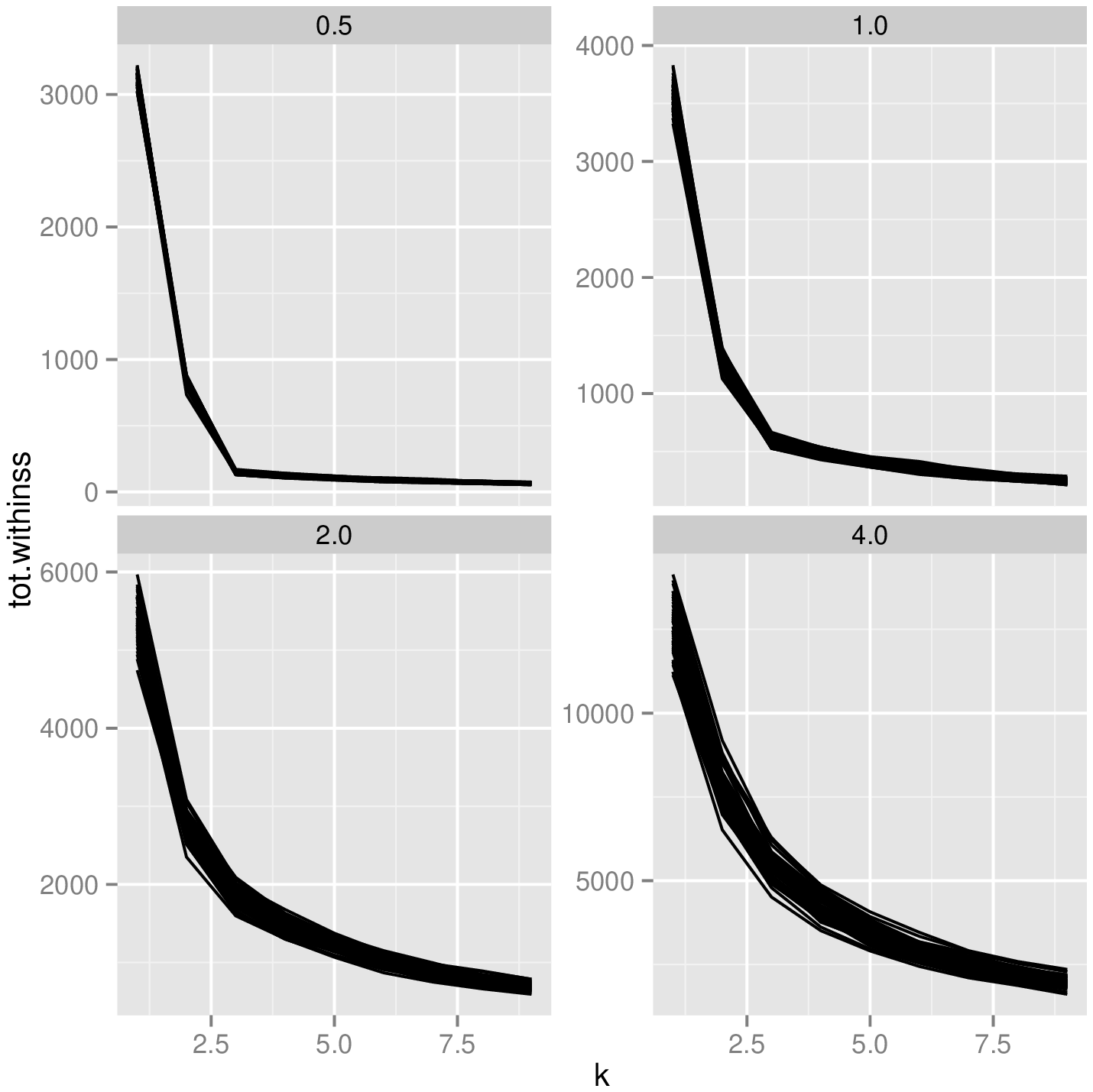} 

\end{knitrout}
\end{center}

We can observe from this that the choice of $k$ based on the total within sum-of-squares profile becomes more difficult as $\sigma$ increases, since the bend at $k=3$ becomes less distinct. We could even measure the cluster purity, and see how the classification accuracy depends on $\sigma$, focusing on the cases where we (correctly) set $k=3$. This requires some processing but can be done entirely in \pkg{dplyr} operations.

\begin{center}
\begin{knitrout}
\definecolor{shadecolor}{rgb}{0.969, 0.969, 0.969}\color{fgcolor}\begin{kframe}
\begin{alltt}
\hlstd{accuracies} \hlkwb{<-} \hlstd{assignments_sd} \hlopt{%>%} \hlkwd{filter}\hlstd{(k} \hlopt{==} \hlnum{3}\hlstd{)} \hlopt{%>%}
    \hlkwd{count}\hlstd{(replication, sd, oracle, .cluster)} \hlopt{%>%}
    \hlkwd{group_by}\hlstd{(replication, sd, .cluster)} \hlopt{%>%}
    \hlkwd{summarize}\hlstd{(}\hlkwc{correct} \hlstd{=} \hlkwd{max}\hlstd{(n),} \hlkwc{total} \hlstd{=} \hlkwd{sum}\hlstd{(n))} \hlopt{%>%}
    \hlkwd{group_by}\hlstd{(replication, sd)} \hlopt{%>%}
    \hlkwd{summarize}\hlstd{(}\hlkwc{purity} \hlstd{=} \hlkwd{sum}\hlstd{(correct)} \hlopt{/} \hlkwd{sum}\hlstd{(total))}
\hlkwd{ggplot}\hlstd{(accuracies,} \hlkwd{aes}\hlstd{(}\hlkwd{factor}\hlstd{(sd), purity))} \hlopt{+} \hlkwd{geom_boxplot}\hlstd{()}
\end{alltt}
\end{kframe}
\includegraphics[width=3in,height=3in]{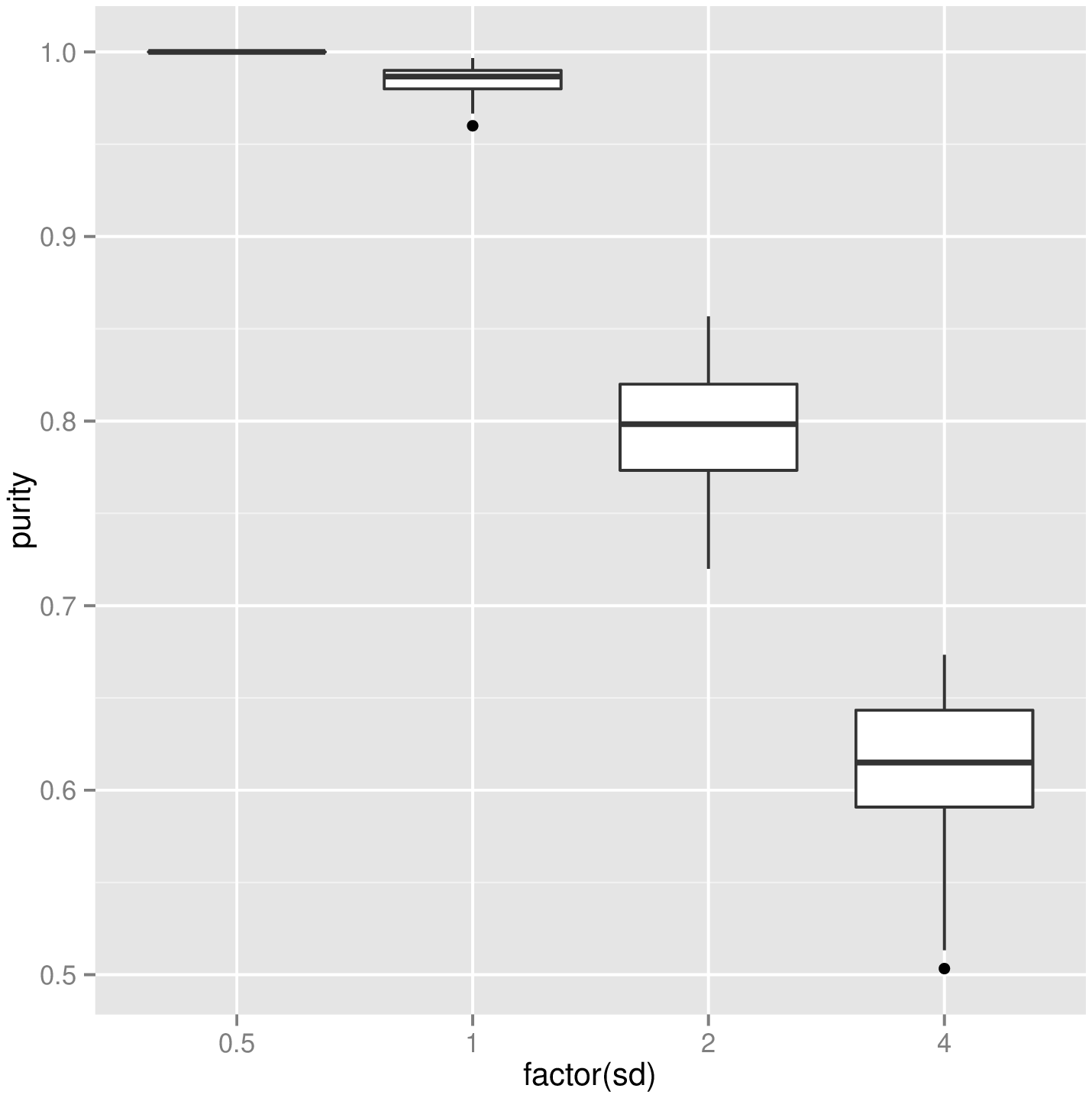} 

\end{knitrout}
\end{center}

We can see that, as one might expect, the classification accuracy decreases on average as the residual standard deviation increases and the clusters get more disperse. These examples demonstrate that combining the models from simulations into a tidied form thus lends itself well to many kinds of exploratory analyses and experiments.

\section{Discussion}

In this paper I introduce the \pkg{broom} package, define the \code{tidy}, \code{augment}, and \code{glance} generics, and describe standards for their behavior. I then provide case studies that illustrate how the tidied outputs that \textbf{broom} creates are useful in a variety of analyses and simulation. The examples listed here are by no means meant to be exhaustive, and indeed make use of only a minority of the tidying methods implemented by \pkg{broom}. Rather, they are meant to suggest the diversity of visualizations and analyses made possible by tidied model outputs.

While \pkg{broom} provides implementations of these generics for many popular R objects, there is no reason that such implementations should be confined to this package. In the future, packages that wish to be output-tidy while retaining the structure of their output object could provide their own \code{tidy}, \code{augment}, and \code{glance} implementations. This would take advantage of each developer's familiarity with his or her software and its goals while offering users a standard tidying language for working with model outputs.

Tidying model outputs is not an exact science, and it is based on a judgment of the kinds of values a data scientist typically wants out of a tidy analysis (for instance, estimates, test statistics, and p-values). Any implementation may lose some of the information that a user wants or keep more information than one needs. It is my hope that data scientists will propose and contribute their own features to expand and improve the functionality of \pkg{broom} and to advance the entire available suite of tidy tools.

\section{Acknowledgments}

I thank Andy Bass, Dmitry Gorenshteyn, and Bob Muenchen for helpful comments and discussions on this manuscript. I thank Hadley Wickham for comments early in \pkg{broom}'s development and for contributing code from \pkg{ggplot2} to the package, and thank Matthieu Gomez and Boris Demeshev for contributions to the software.

This work was supported in part by NIH (R01 HG002913).

\bibliographystyle{abbrvnat}
\bibliography{RJreferences}


\end{document}